\documentclass[twoside]{IEEEtran}
\usepackage[utf8]{inputenc}

\usepackage{amsmath,amssymb} \interdisplaylinepenalty=2500
\usepackage{cite}
\usepackage{graphicx}

\usepackage{algorithm}
\usepackage{algorithmic}
\usepackage{hyperref}

\newcommand{\vnr}{\text{VNR}}
\newcommand{\wer}{\text{WER}}

\newtheorem{theorem}{Theorem}
\newtheorem{proposition}{Proposition}
\newtheorem{corollary}{Corollary}
\newtheorem{lemma}{Lemma}
\newtheorem{definition}{Definition}
\newenvironment{remark}{\textit{Remark: }}{}
\newtheorem{myexample}{Example}
\newenvironment{example}{\myexample}{\qed\endmyexample}
\def\qed{\endIEEEproof}

\newcommand{\mat}[1]{\begin{bmatrix} #1 \end{bmatrix}}

\makeatletter
\renewcommand*\env@matrix[1][*\c@MaxMatrixCols c]{	\hskip -\arraycolsep
	\let\@ifnextchar\new@ifnextchar
	\array{#1}}
\makeatother

\DeclareMathOperator*{\argmax}{argmax}

\newcommand{\bzero}{\mathbf{0}}
\newcommand{\ba}{\mathbf{a}}
\newcommand{\bb}{\mathbf{b}}
\newcommand{\bc}{\mathbf{c}}

\newcommand{\bh}{\mathbf{h}}

\newcommand{\bp}{\mathbf{p}}

\newcommand{\br}{\mathbf{r}}
\newcommand{\bs}{\mathbf{s}}

\newcommand{\bu}{\mathbf{u}}
\newcommand{\bv}{\mathbf{v}}

\newcommand{\bx}{\mathbf{x}}
\newcommand{\by}{\mathbf{y}}
\newcommand{\bz}{\mathbf{z}}

\newcommand{\bB}{\mathbf{B}}

\newcommand{\bD}{\mathbf{D}}
\newcommand{\bE}{\mathbf{E}}
\newcommand{\bF}{\mathbf{F}}
\newcommand{\bG}{\mathbf{G}}
\newcommand{\bH}{\mathbf{H}}
\newcommand{\bI}{\mathbf{I}}

\newcommand{\bL}{\mathbf{L}}

\newcommand{\bR}{\mathbf{R}}

\newcommand{\bT}{\mathbf{T}}

\newcommand{\blambda}{\boldsymbol{\lambda}}

\newcommand{\calA}{\mathcal{A}}

\newcommand{\calC}{\mathcal{C}}
\newcommand{\calD}{\mathcal{D}}

\newcommand{\calI}{\mathcal{I}}
\newcommand{\calJ}{\mathcal{J}}
\newcommand{\calK}{\mathcal{K}}

\newcommand{\calQ}{\mathcal{Q}}
\newcommand{\calR}{\mathcal{R}}

\newcommand{\calV}{\mathcal{V}}

\newcommand{\ZZ}{\mathbb{Z}}

\newcommand{\RR}{\mathbb{R}}
\newcommand{\FF}{\mathbb{F}}

\newcommand{\transpose}{\mathrm{T}}
\newcommand{\T}{\mathrm{T}}

\begin{document}

\title{Multilevel LDPC Lattices with Efficient Encoding and Decoding and a Generalization of Construction~D$'$}

\author{Paulo~Ricardo~Branco~da~Silva
        and Danilo~Silva,~\IEEEmembership{Member,~IEEE}%
\thanks{Manuscript received December 12, 2017; revised July 11, 2018 and November 14, 2018; accepted November 18, 2018. Date of publication MMMM DD, YYYY; date of current version MMMM DD, YYYY. The work of P. R. Branco da Silva was supported by CAPES-Brazil under Grant 1587757. The work of D.~Silva was supported in part by CNPq-Brazil under Grants 429097/2016-6 and 310343/2016-0. This paper was presented in part at the XXXV Simp\'osio Brasileiro de Telecomunica\c{c}\~oes e Processamento de Sinais, S\~ao Pedro, Brazil, September 2017, and in part at the IEEE International Symposium on Information Theory, Vail, CO, June 2018. The associate editor coordinating the review of this paper and approving it for publication was F.~Oggier. \textit{(Corresponding author: Danilo Silva)}.}%
\thanks{P. R. Branco da Silva and D. Silva are with the Department of Electrical and Eletronic Engineering, Federal University of Santa Catarina, Florianopolis-SC, Brazil (e-mails: paulo.branco@posgrad.ufsc.br; danilo.silva@ufsc.br).}%
\thanks{Color versions of one or more of the figures in this paper are available
online at http://ieeexplore.ieee.org.}%
\thanks{Digital Object Identifier XXXXX}%
}

\markboth{IEEE Transactions on Information Theory, to appear}{Branco da Silva and Silva: Multilevel LDPC Lattices with Efficient Encoding and Decoding}

\maketitle

\begin{abstract}
Lattice codes are elegant and powerful structures that not only can achieve the capacity of the AWGN channel but are also a key ingredient to many multiterminal schemes that exploit linearity properties. However, constructing lattice codes that can realize these benefits with low complexity is still a challenging problem. In this paper, efficient encoding and decoding algorithms are proposed for multilevel binary LDPC lattices constructed via Construction~D$'$ whose complexity is linear in the total number of coded bits. Moreover, a generalization of Construction~D$'$ is proposed that relaxes some of the nesting constraints on the parity-check matrices of the component codes, leading to a simpler and improved design. Based on this construction, low-complexity multilevel LDPC lattices are designed whose performance under multistage decoding is comparable to that of polar lattices and close to that of low-density lattice codes (LDLC) on the power-unconstrained AWGN channel.
\end{abstract}


\section{Introduction}

\IEEEPARstart{L}{attice} codes, the analogue of linear codes in the Euclidean space, have attracted an increasing amount of attention in recent years. 
Their rich structure not only provides an elegant and powerful solution to achieving the capacity of the additive white Gaussian noise (AWGN) channel, but is also a key ingredient to many multiterminal information theory schemes that exploit linearity properties \cite{Zamir.2014.Lattice-Coding-Signals}. However, despite their theoretical success, constructing lattice codes that can realize these benefits with low complexity is still a challenging problem.

A promising direction is the use of lattices with a low-density parity-check structure, which admit a belief propagation decoder with complexity linear in the lattice dimension. While several constructions of this form have been proposed recently and shown to achieve remarkable performance \cite{Sommer.etal.2008.Low-density-Lattice-Codes,Pietro.etal.2012.Integer-Low-density-Lattices,Boutros.etal.2014.Generalized-Low-density-(GLD)}, they all suffer from the need to perform operations over a large field (either $\FF_p$ \cite{Pietro.etal.2012.Integer-Low-density-Lattices,Boutros.etal.2014.Generalized-Low-density-(GLD)} or $\RR$ \cite{Sommer.etal.2008.Low-density-Lattice-Codes}), whose complexity per bit is much higher than that of binary codes.

Alternative lattice constructions that leverage
the use of binary codes are the multilevel Construction~D and Construction~D$'$ \cite{Barnes.Sloane.1983.New-Lattice-Packings,Conway.Sloane.1999.Sphere-Packings-Lattices}, which rely on a family of $L$ nested binary linear codes used in conjunction with $2^L$-PAM modulation. The former (latter) describes a lattice through the generator (parity-check) matrices of the nested component codes.
When used together with multistage decoding (MSD) \cite{Wachsmann.etal.1999.Multilevel-Codes:-Theoretical,Forney.etal.2000.Sphere-bound-achieving-Coset-Codes}, each component code can be individually encoded and decoded over the binary field, leading to a significant complexity reduction.

Construction~D$'$ lattices based on binary low-density parity-check (LDPC) codes, referred to as LDPC lattices, were originally introduced in \cite{Sadeghi.etal.2006.Low-density-Parity-check-Lattices:} and subsequently studied in \cite{Choi.etal.2008.Iterative-Decoding-Low-Density,Baik.Chung.2008.Irregular-Low-density-Parity-check,Safarnejad.Sadeghi.2012.FFT-Based-Sum-Product}. To the best of our knowledge, no further papers have been published on the theory or applications of strictly multilevel ($L \geq 2$) Construction~D$'$ LDPC lattices. This apparent lack of interest may be partly explained by three major challenges of Construction~D$'$ which have so far remained unsolved:
\begin{itemize}
\item \textbf{Lack of efficient encoding:} The complexity of encoding an LDPC lattice is quadratic in the lattice dimension if done naively using the lattice generator matrix. In order to exploit low-complexity systematic encoding methods for LDPC codes, one should be able to encode each component code individually using its parity-check matrix. However, in the definition of Construction~D$'$, individual levels are coupled through non-binary modular equations in a nontrivial way, making it unclear how to perform individual encoding.
\item \textbf{Lack of efficient multistage decoding:} In principle, multistage decoding requires the influence of all past levels to be removed---by re-encoding and subtracting from the received vector---before the current level is decoded. However, when using Construction~D$'$, it is unclear how a level influences the subsequent ones, so that re-encoding can be performed. Moreover, due to such dependence on re-encoding, an efficient encoding method would be required for multistage decoding to be efficient.
\item \textbf{Performance gap:} Even if we ignore the issue of low-complexity decoding, experimental results reveal that the finite-length performance of existing Construction~D$'$ LDPC lattices is typically much inferior to that of Construction~D lattices, contradicting what one might expect from the excellent performance of LDPC codes. Part of the reason may be that Construction~D$'$ requires not only the component codes, but also their parity-check matrices, to be nested (i.e., one matrix must contain the other as a submatrix), a stringent constraint that may degrade the overall performance of the resulting lattice.
\end{itemize}

In this paper, these three challenges are finally solved starting from a reinterpretation of Construction~D$'$. 

Our main technical contributions are summarized as follows:
\begin{itemize}
\item We present an alternative description of Construction~D$'$ which enables sequential encoding of the component codes. The contribution of past levels on the current level is subsumed by a syndrome vector that the current codeword must satisfy. Apart from that, the actual encoding of a level is done entirely over the binary field. Moreover, as the syndrome vector can be computed directly from parity-check equations, the need for explicit re-encoding is avoided for multistage decoding.
\item We show how existing linear-time algorithms for encoding and decoding LDPC codes can be adapted to handle cosets of LDPC codes without increasing the order of complexity. It follows that multilevel LDPC lattices can always be decoded with complexity $O(L n)$ and can be encoded also with complexity $O(Ln)$ if the component codes admit linear-time systematic encoding, where $n$ is the lattice dimension. 
Moreover, encoding and decoding can be performed with off-the-shelf binary LDPC encoders and decoders.
\item We propose a generalization of Construction~D$'$ that relaxes the nesting constraints on the parity-check matrices of the component codes (which themselves must remain nested). 
This new construction significantly enlarges the design space for good component codes,
enabling the design of multilevel LDPC lattices with much better performance. This contribution may also be of independent interest from a mathematical perspective.
\item We propose an efficient method to construct the parity-check matrices of the remaining component codes given only the parity-check matrix of the highest-rate LDPC code, while respecting the conditions of the generalized Construction~D$'$. We also present a variation of this basic method aimed at maximizing girth and another that enables linear-time encoding.
\item We present examples of two-level LDPC lattices, with encoding and decoding complexity linear in the total number of coded bits, that 
achieve performance comparable to state-of-the-art Construction~D lattices \cite{Yan.etal.2014.Construction-Capacity-achieving-Lattice}, closing a long-standing gap.
\end{itemize}

It is worth mentioning that, although we have opted to use binary codes for simplicity and computational efficiency, our results can be straightforwardly generalized to a general prime $p$, as well as to Complex Construction~D$'$ lattices over a complex ring \cite{Conway.Sloane.1999.Sphere-Packings-Lattices,Forney.1988.Coset-Codes---Part-I:,Feng.etal.2011.Lattice-Network-Coding-1}.

\subsection{Related Work}

A lattice code consists of the intersection of a lattice (a discrete additive subgroup of the Euclidean space) and a bounded region, also called a shaping region. While it is well-known that lattice codes can achieve the capacity of the AWGN channel (see \cite{Urbanke.Rimoldi.1998.Lattice-Codes-Can} and references therein), renewed interest in the topic can be traced to the seminal paper by Erez and Zamir \cite{Erez.Zamir.2004.Achieving-1/2-Log}, who showed that capacity can be achieved by nested lattice codes with lattice decoding (a suboptimal decoding approach which effectively ignores the shaping region). Since then, several applications of lattice codes to multiterminal information theory have been proposed based on their results, including: distributed source coding \cite{Krithivasan.Pradhan.2009.Lattices-Distributed-Source}, physical-layer security \cite{Ling.etal.2014.Semantically-Secure-Lattice}, and communication over Gaussian networks \cite{Zamir.2014.Lattice-Coding-Signals}---in particular, lattice codes are essential to the compute-and-forward strategy for relay networks \cite{Nazer.Gastpar.2011.Compute-and-Forward:-Harnessing-Interference} and to integer-forcing methods for MIMO channels \cite{Zhan.etal.2014.Integer-forcing-Linear-Receivers}.

The main problem in the design of a lattice code is arguably the design of the underlying lattice, which must be good at rejecting noise. This problem can be formalized with the power-unconstrained AWGN channel model introduced by Poltyrev \cite{Poltyrev.1994.Coding-Restrictions-AWGN}, leading to the notion of AWGN-goodness 
as a necessary condition to achieve capacity. As a consequence, much of the literature on the topic, as well as the current paper, is focused on the unconstrained lattice design problem. Recent work has shown that even the use of a fixed shaping region is not essential to achieve capacity under lattice decoding, if signal points are selected with a nonuniform (discrete Gaussian) distribution \cite{Ling.Belfiore.2014.Achieving-AWGN-Channel,Campello.Dadush.2017.AWGN-Goodness-Enough:-Capacity-Achieving}. In principle, the lattices designed in the present paper can be combined with a variety of shaping methods, such as trellis shaping \cite{Forney.1992.Trellis-Shaping} or probabilistic shaping \cite{Ling.Belfiore.2014.Achieving-AWGN-Channel,Campello.Dadush.2017.AWGN-Goodness-Enough:-Capacity-Achieving}, which are however outside the scope of this work.

A popular way of constructing lattices, which exploits the power of linear codes, is the so-called Construction A \cite{Conway.Sloane.1999.Sphere-Packings-Lattices}. The basic (real-valued) version of this construction relies on a single $p$-ary linear code, where $p$ is a prime, which is repeated across the Euclidean space (at multiples of $p$ along each coordinate) to produce an infinite constellation. This construction is the source of most proofs of achievable rates (using asymptotically long random codes) \cite{Erez.Zamir.2004.Achieving-1/2-Log,Zamir.2014.Lattice-Coding-Signals}, but is also shown to produce lattices with excellent finite-length performance and complexity linear in the lattice dimension, provided that $p$ is sufficiently large. This is the case with Generalized Low-Density (GLD) lattices \cite{Boutros.etal.2014.Generalized-Low-density-(GLD),Pietro.etal.2015.Non-binary-GLD-Codes} and Low-Density Construction A (LDA) lattices \cite{Pietro.etal.2012.Integer-Low-density-Lattices,Pietro.etal.2013.New-Results-Construction,Pietro.etal.2018.LDA-Lattices-Dithering}, which are shown to be AWGN-good. However, decoding a $p$-ary code, for large $p$, is much more complex than decoding a similarly structured binary code. For instance, the belief propagation decoder in \cite{Pietro.etal.2012.Integer-Low-density-Lattices} has complexity $O(p^2n)$, i.e., the complexity is exponential in the bit depth $\log_2 p$. 

Another approach is to construct lattices that are designed and decoded directly in Euclidean space, such as low-density lattice codes (LDLC) \cite{Sommer.etal.2008.Low-density-Lattice-Codes}. However, since the decoder now has to process continuous functions \cite{Sommer.etal.2008.Low-density-Lattice-Codes,Hernandez.Kurkoski.2016.Three/two-Gaussian-Parametric}, the decoding complexity is typically even higher than that of Construction~A lattices.

Multilevel lattice constructions based on binary codes are potentially harder to design, but have the promise of complexity that scales linearly with the number of levels. Moreover, they are known to be AWGN-good under multistage decoding \cite{Forney.etal.2000.Sphere-bound-achieving-Coset-Codes}. Construction~D has been used in \cite{Sakzad.etal.2010.Construction-Turbo-Lattices} to produce turbo lattices and in \cite{Yan.etal.2014.Construction-Capacity-achieving-Lattice} to construct polar lattices; the latter are shown to be AWGN-good with encoding and decoding complexity $O(L n \log n)$. Construction~D has also been used in \cite{Vem.etal.2014.Multilevel-Lattices-Based} to construct spatially-coupled LDPC lattices, which were shown to be AWGN-good under multistage belief propagation decoding. However, both the encoding and the cancellation step that has to be performed at each decoding stage rely on the generator matrices of the component LDPC codes, which are generally dense, leading to an overall high complexity.

Construction~D$'$ has been used in \cite{Sadeghi.etal.2006.Low-density-Parity-check-Lattices:,Choi.etal.2008.Iterative-Decoding-Low-Density,Baik.Chung.2008.Irregular-Low-density-Parity-check,Safarnejad.Sadeghi.2012.FFT-Based-Sum-Product} to construct multilevel LDPC lattices; however, these works consider only joint decoding\footnote{The multistage decoder mentioned in \cite{Baik.Chung.2008.Irregular-Low-density-Parity-check} is unsuitable for Construction~D$'$ since it relies on independent encoding of levels through the Code Formula \cite{Forney.1988.Coset-Codes---Part-I:}, which does not generally produce lattices \cite{Kositwattanarerk.Oggier.2014.Connections-Construction-Related}.} of the component codes, whose complexity is exponential in $L$, while the encoding complexity is not addressed. Both issues can be avoided using a single-level LDPC lattice, as done in \cite{Sadeghi.Sakzad.2013.Performance-1-level-LDPC,Khodaiemehr.etal.2015.One-level-LDPC-Lattice,Khodaiemehr.etal.2016.Construction-Full-diversity-1-level,Khodaiemehr.etal.2017.LDPC-Lattice-Codes,Khodaiemehr.etal.2017.Practical-Encoder-Decoder}, allowing the use of conventional encoding and decoding methods for LDPC codes. However, in this case the construction reduces to Construction~A, which is known to result in a poor performance for $p=2$. More precisely, since all the higher levels are uncoded, the performance in terms of word error rate quickly degrades as the block length increases and does not improve sharply as the noise level decreases.

\subsection{Organization}

The remainder of this paper is organized as follows. Section~\ref{sec:preliminaries} reviews basic concepts on lattices. In Section~\ref{sec:alternative_descriptions}, we present a sequential description of Construction~D$'$, which is then used in Section~\ref{sec:codec} to derive efficient multilevel encoding and multistage decoding for Construction~D$'$ LDPC lattices. In Section~\ref{sec:generalized}, a generalization of Construction~D$'$ is proposed that reduces the nesting constraints to a minimum while still satisfying the requirements for sequential encoding. In Section~\ref{sec:generalized}, we propose and extend an efficient method to construct Generalized Construction~D$'$ LDPC lattices, which we call check splitting. Section~\ref{sec:results} presents design examples and corresponding simulation results and Section~\ref{sec:conclusion} presents our conclusions.

\section{Preliminaries}
\label{sec:preliminaries}

In the following, we describe our notation and review basic concepts on lattices, in particular, Construction D$'$. For further details, we refer the reader to \cite{Zamir.2014.Lattice-Coding-Signals,Conway.Sloane.1999.Sphere-Packings-Lattices,Forney.etal.2000.Sphere-bound-achieving-Coset-Codes}.

\subsection{Notation}
\label{ssec:notation}

Let $\bzero$ denote the all-zero vector, with appropriate length implied by the context. If $\calA$ is a set, then $\calA^n$ and $\calA^{m \times n}$ denote the set of length-$n$ vectors and $m \times n$ matrices, respectively, with entries in $\calA$. Let $\FF_2$ be the finite field of size $2$ and let $\varphi: \ZZ \to \ZZ/2\ZZ \cong \FF_2$ be the natural reduction homomorphism, extended to vectors and matrices in a component-wise fashion.

We follow common convention for the notation ``$\bmod \; m$'' when $m$ is an integer.
For any $a,b \in \ZZ$, we use the modular congruence $a \equiv b \pmod{m}$ to denote that $a-b$ is divisible by $m$. For any $x \in \RR$, we define $x \bmod m$ as the unique $r \in [0,m)$ such that $x = r + qm$, for some $q \in \ZZ$. These notations are again extended to vectors and matrices in a component-wise fashion.

In this paper, we treat binary linear codes and associated parity-check matrices as having entries in $\{0,1\} \subseteq \ZZ$, rather than in $\FF_2$. This approach significantly simplifies notation when dealing with lattices. The algebraic properties of a linear code are recovered by using modular equations or the explicit mapping to $\FF_2$. For instance, if $\bH \in \{0,1\}^{m \times n}$, then a binary linear code $\calC \subseteq \{0,1\}^n$ is defined by the parity-check matrix $\bH$ as
\[
\calC = \{\bx \in \{0,1\}^n: \bH \bx^T \equiv \bzero \pmod{2}\}.
\]
The dimension of $\calC$ as a linear code is the dimension of the subspace $\varphi(\calC) \subseteq \FF_2^n$ or, equivalently, the dimension of the null space of $\varphi(\bH) \in \FF_2^{m \times n}$.

\subsection{Lattices}
\label{ssec:lattices}

A lattice $\Lambda \subseteq \RR^n$ is a discrete subgroup of $\RR^n$. This implies that $\Lambda$ is closed under integer linear combinations and may be expressed as $\Lambda = \{\bu \bG, \; \bu \in \ZZ^n \}$, where $\bG \in \RR^{n \times n}$ is a generator matrix. 

A fundamental region of $\Lambda$ is a set $\calR_{\Lambda} \subseteq \RR^n$ 
such that any $\bx \in \RR^n$ can be \textit{uniquely} expressed as $\bx = \blambda + \br$, where $\blambda \in \Lambda$ and $\br \in \calR_{\Lambda}$. Every fundamental region has the same volume, which is denoted by $V(\Lambda)$. A fundamental region $\calR_{\Lambda}$ defines a quantizer $Q_\Lambda: \RR^n \to \Lambda$ and a modulo-$\Lambda$ operation $\RR^n \to \calR_\Lambda$ as $Q_\Lambda(\bx) = \blambda$ and $\bx \bmod \Lambda = \br$, respectively, where $\bx = \blambda + \br$. In particular, the Voronoi region of $\Lambda$ (around the origin) is the set $\calV_\Lambda$ of points that are closer to $\bzero$ than to any other lattice point, with ties decided arbitrarily but such that $\calV_\Lambda$ is a fundamental region.

A sublattice $\Lambda' \subseteq \Lambda$ is a subset of $\Lambda$ which is itself a lattice. If $\Lambda$ and $\Lambda' \subseteq \Lambda$ are lattices, then $\calC = \Lambda \cap \calR_{\Lambda'}$ is said to be a nested lattice code. Note that $|\calC| = V(\Lambda')/V(\Lambda)$.

\subsection{Transmission Without a Power Constraint}
\label{ssec:poltyrev}

In the design of lattice codes, one is often interested in addressing the coding problem separately from the shaping problem. This leads to the so-called (power-)unconstrained AWGN channel studied by Poltyrev \cite{Poltyrev.1994.Coding-Restrictions-AWGN}, over which any lattice point may be transmitted without restrictions. In that case, the main performance metric for a lattice is its probability of decoding error for a given density of lattice points.

More precisely, let $\Lambda \subseteq \RR^n$ be a lattice. The channel output is given by $\by = \bx + \bz$, where $\bx \in \Lambda$ is the transmitted vector and $\bz \in \RR^n$ is a white Gaussian noise vector with variance $\sigma^2$ per component. Decoding is performed by quantizing $\by$ to the nearest lattice point $\hat{\bx} = \calQ_\Lambda(\by)$. The probability of error, denoted by $P_e(\Lambda,\sigma^2)$, is the probability that $\bz$ falls outside~$\calV_\Lambda$.

Given that the density of $\Lambda$ is inversely proportional to $V(\Lambda)$, it is convenient to define the volume-to-noise ratio (VNR) as\footnote{The VNR is also commonly defined \cite{Zamir.2014.Lattice-Coding-Signals} without the term $2\pi e$ in the denominator.}
\begin{equation}
\label{eq:volume_to_noise_ratio}
\gamma_{\Lambda} (\sigma) \triangleq \frac{V (\Lambda)^{2/n} }{ 2 \pi e \sigma^2 }
\end{equation}
which gives a measure of the density of $\Lambda$ relative to the noise level.

It is well-known \cite{Zamir.2014.Lattice-Coding-Signals,Forney.etal.2000.Sphere-bound-achieving-Coset-Codes} that, if $P_e (\Lambda, \sigma^2) \approx 0$, then $\gamma_{\Lambda} (\sigma)$ is necessarily greater than $1$, or $0$~dB. This fundamental limit 
is known as the Poltyrev limit or the sphere bound. On the other hand, for all $\sigma^2>0$ and all $P_e > 0$, there exists a sequence of $n$-dimensional lattices with $P_e (\Lambda, \sigma^2) \leq P_e$ such that $\lim_{n\to \infty} \gamma_{\Lambda} (\sigma) = 1$. Lattices with this property are said to be \textit{good} for AWGN coding.

In practice, nearest-neighbor lattice decoding may not be feasible to implement, so a suboptimal decoder $\calD: \RR^n \to \Lambda$ may be used instead. In this case, one should refer to the probability of error of the pair $(\Lambda, \calD)$, i.e., $P_e((\Lambda,\calD),\sigma^2)$. For simplicity, we keep the notation $P_e(\Lambda,\sigma^2)$ when the decoder is clear from the context.

\subsubsection{Forney-Trott-Chung's Two-Stage Decoding}
\label{ssec:forney}

A practical way to approach the Poltyrev problem, which often simplifies the decoding and the code design, is the two-level partition proposed in \cite{Forney.etal.2000.Sphere-bound-achieving-Coset-Codes}. Given a lattice $\Lambda \subseteq \RR^n$, a sublattice $\Lambda' \subseteq \Lambda$ is chosen such that the modulo-$\Lambda'$ operation (as well as the enumeration of elements of $\Lambda'$) is easy to implement. In this manner, $\Lambda$ can be partitioned as $\Lambda = \calC + \Lambda'$, where $\calC = \Lambda \cap \calR_{\Lambda'}$ is a lattice code and $\calR_{\Lambda'}$ is a fundamental region defining the modulo-$\Lambda'$ operation.

In order to transmit $\bx \in \Lambda$, vectors $\bc \in \calC$ and $\blambda' \in \Lambda'$ are chosen independently. The point $\bx$ is transmitted as the sum of these two components, i.e., $\bx = \bc + \blambda'$. Let $\by = \bx + \bz$ be the channel output, as described above. Decoding from $\by$ is as follows. First,
\begin{equation}\label{eq:equiv-channel-forney}
\br = \by \bmod \Lambda' = \bc + \bz \bmod \Lambda'
\end{equation}
is computed, eliminating the influence of $\blambda'$. For instance, if $\Lambda' = q\ZZ^n$, with $\calR_{\Lambda'} = [0,q)^n$, then the modulo-$\Lambda'$ operation can be easily implemented by component-wise modulo-$q$ reduction over $\RR$ (see Section~\ref{ssec:notation}). Next, a decoder for code $\calC$ is applied on the modulo-$\Lambda'$ equivalent channel described above, from which the estimate $\hat{\bc} \in \calC$ is obtained. Finally, $\hat{\bc}$ is subtracted from $\by$, resulting in
\begin{equation}
\by' = \by - \hat{\bc} = (\bc - \hat{\bc}) + \blambda' + \bz,
\end{equation} 
and a decoder for $\Lambda'$ is applied, obtaining the estimate $\hat{\blambda'} \in \Lambda'$ and, consequently, $\hat{\bx} = \hat{\bc} + \hat{\blambda'}$.

It follows by the union bound that \begin{equation}
\label{eq:union_bound}
P_e (\Lambda,\sigma^2) \leq P_e (\calC,\Lambda') + P_e (\Lambda',\sigma^2)
\end{equation}
where $P_e (\calC,\Lambda')$ is the probability of error for the code $\calC$ when used on the channel in \eqref{eq:equiv-channel-forney}. Note that, when $\Lambda' = q\ZZ^n$, nearest-neighbor lattice decoding can be easily implemented and $P_e(\Lambda',\sigma^2)$ can be computed exactly, i.e.,
\begin{equation}
\label{eq:an-Pe-4ZZ}
P_e(q\ZZ^n,\sigma^2) = 1-\left(1-2Q\left(\frac{q}{2\sigma}\right)\right)^n
\end{equation}
where $Q(x) = (1/\sqrt{2\pi})\int_{x}^{\infty}e^{-u^2/2}du$ is the Q-function.

\subsection{Construction D$'$} 
\label{ssec:Dprime}

Let $\calC_0 \subseteq \calC_1 \subseteq \cdots \subseteq \calC_{L-1} \subseteq \{0,1\}^n$ be a family of nested binary linear codes. For $\ell = 0,\ldots,L-1$, let $k_\ell$ be the dimension of $\calC_\ell$, let $R_\ell = k_\ell/n$, and let $m_{\ell} = n - k_{\ell}$. For convenience, define $m_L=0$. Clearly,
\begin{equation*}
m_0 \geq m_1 \geq \cdots \geq m_{L-1} \geq m_L.
\end{equation*}
Let $\bh_1,\ldots, \bh_{m_0} \in \{0,1\}^n$ 
be such that
\begin{equation}\label{eq:Hmatrix}
\bH_\ell = \mat{\bh_1 \\ \vdots \\ \bh_{m_{\ell}}}
\end{equation}
is a parity-check matrix for~$\calC_{\ell}$, for $\ell = 0,\ldots,L-1$. 
An $L$-level Construction D$'$ lattice \cite{Conway.Sloane.1999.Sphere-Packings-Lattices} is defined as
\begin{align}
\label{eq:parity_congruences}
\Lambda = \{ \bv \in \ZZ^n: {}&\bh_j \bv^\T \equiv 0 \pmod{2^{\ell+1}}, \nonumber \\ &m_{\ell + 1} < j \leq m_{\ell},\; 0\leq \ell \leq L-1 \}.
\end{align}

We can also express $\Lambda = \calC + 2^L \ZZ^n$, where  
\begin{equation}\label{eq:lattice-code}
\calC = \Lambda \cap [0,2^L)^n
\end{equation}
is a lattice code. In particular, $V(\Lambda) = 2^{n(L-R)}$, where 
\begin{equation}
R = R(\calC) \triangleq \frac{1}{n}\log_2|\calC|.
\end{equation}

If matrices $\bH_0,\ldots,\bH_{L-1}$ are sparse, i.e., if $\calC_0,\ldots,\calC_{L-1}$ are LDPC codes, then $\Lambda$ is said to be an $L$-level LDPC lattice~\cite{Sadeghi.etal.2006.Low-density-Parity-check-Lattices:}. 

It is worth emphasizing that the number of levels in the construction, $L$, refers to the number of \textit{coded} levels---there is always an additional uncoded level ($\ell=L$) corresponding to the lattice $2^L \ZZ^n$, which tiles the lattice code $\calC$ into an infinite constellation. Note that, for $L=1$, the construction is ``degenerate'' in the sense that it reduces to Construction~A \cite{Conway.Sloane.1999.Sphere-Packings-Lattices}. Thus, strictly multilevel Construction~D$'$ lattices require $L \geq 2$. For the remainder of the paper, we assume $L \geq 2$ when referring to Construction D$'$.

\begin{remark}
When an $L$-level Construction D$'$ lattice $\Lambda \subseteq \ZZ^n$ is used for the Poltyrev channel under the approach of Section~\ref{ssec:forney}, we naturally choose $\Lambda' = 2^L \ZZ^n$ and $\calR_{\Lambda'} = [0,2^L)^n$, so that the $\bmod~\Lambda'$ operation becomes simply $\bmod~2^L$ over~$\RR$. In this case, decoding of $\Lambda$ essentially reduces to decoding of the lattice code $\calC$ over the mod-$2^L$ channel $\by = \bc + \bz \bmod 2^L$, where $\bc \in \calC$.
\end{remark}

\section{A Sequential Description of Construction D$'$} 
\label{sec:alternative_descriptions}

In this section, an alternative description of Construction~D$'$ is proposed that enables sequential multilevel encoding based on the original component codes. We start by rewriting \eqref{eq:parity_congruences} in matrix form, which will also be useful for the results in Section~\ref{sec:generalized}.

\medskip
\begin{proposition}\label{th:matrix_description}
Let $\Lambda$ be a lattice defined by \eqref{eq:parity_congruences}.
Then
\begin{equation}
\label{eq:lambda_matrix_definition}
\Lambda = \left\{ \bv \in \ZZ^n: \bH_\ell \bv^\T \equiv \bzero \pmod{2^{\ell+1}},\; 0\leq \ell \leq L-1 \right\}.
\end{equation}
\end{proposition}
\begin{IEEEproof}
The proof is straightforward and is omitted.
\end{IEEEproof}

\medskip
\begin{lemma}
\label{lem:matrix-nesting}
If $\bH_0,\ldots,\bH_{L-1}$ are matrices satisfying \eqref{eq:Hmatrix}, then these matrices have the property that, for all $\bv \in \ZZ^n$ and all $1 \leq \ell \leq L-1$, 
\begin{equation}
\label{eq:lemma1}
\bH_{\ell-1} \bv^T \equiv \bzero \pmod{2^\ell} \, \implies \, \bH_{\ell} \bv^T \equiv \bzero \pmod{2^\ell}.
\end{equation}
\end{lemma}
\begin{IEEEproof}
The proof follows immediately since, by definition, $\bH_{\ell}$ is a submatrix of $\bH_{\ell-1}$.
\end{IEEEproof}
\medskip

We state the following result with slight generality since this will be useful later on. For convenience, define an empty summation as the all-zero vector with the appropriate size.

\medskip
\newcommand{\calCseq}{\calC_{\text{seq}}}
\begin{theorem}[Sequential Encoding]
\label{th:sequential_encoding}
Let $\calC = \Lambda \cap [0,2^L)^n$ be a lattice code carved from a lattice $\Lambda$ defined by \eqref{eq:lambda_matrix_definition}. Suppose that the corresponding matrices $\bH_0,\ldots,\bH_{L-1}$ have the property described in Lemma~\ref{lem:matrix-nesting} and are such that
$\varphi(\bH_\ell) \in \FF_2^{m_\ell \times n}$, $\ell=0,\ldots,L-1$,
are full-rank.
Consider the following procedure:
\begin{enumerate}
\item Sequentially, for $\ell = 0,1,\ldots,L-1$, choose some vector $\bc_\ell \in~\calC_\ell(\bs_\ell)$, where
\begin{equation}
\label{eq:binarycongruence}
\calC_\ell(\bs_\ell) \triangleq \left\lbrace \bv \in \{0,1\}^n: \bH_{\ell} \bv^\T \equiv \bs_{\ell} \pmod{2} \right\rbrace
\end{equation}
is a coset of the linear code $\calC_\ell = \calC_\ell(\bzero)$
and $\bs_{\ell} \in \{0,1\}^{m_\ell}$ is such that
\begin{equation}
\label{eq:syndrome}
\bs_{\ell} \equiv  
\frac{ -\bH_{\ell} \sum_{i=0}^{\ell-1} 2^{i} \bc_{i}^\T}{2^{\ell}} \pmod{2}.
\end{equation}
\item Finally, compute $\bc = \sum_{\ell=0}^{L-1} 2^{\ell} \bc_{\ell}$.
\end{enumerate}

 The procedure described above is well-defined. Moreover, let $\calCseq$ be the set of all possible vectors $\bc \in \ZZ^n$ produced by this procedure. Then $\calCseq = \calC$.
\end{theorem}
\begin{IEEEproof}
First, we prove that the procedure is well-defined, i.e., that 
\begin{equation}\label{eq:proof-well-defined-1}
\bH_\ell \sum_{i = 0}^{\ell-1} 2^i \bc_i^\T \equiv \bzero \pmod{2^{\ell}}
\end{equation}
for $\ell > 0$, so that $\bs_\ell$ can always be computed. Note that a solution for $\bc_{\ell}$ always exists, since $\bH_\ell$ is full-rank modulo~2. We proceed by induction. The base case $\ell = 0$ is already established by definition, since $\bs_0 = \bzero$. Let $\ell > 0$ and suppose that $\bc_0,\ldots,\bc_{\ell-1}$ and $\bs_{\ell-1}$ have been computed. Thus,
\[
2^{\ell-1} \bs_{\ell-1} + \bH_{\ell-1}\sum_{i=0}^{\ell-2} 2^i \bc_i^\T = 2^{\ell} \ba
\]
for some $\ba \in \ZZ^{m_{\ell-1}}$. But $\bH_{\ell-1} \bc_{\ell-1}^T \equiv \bs_{\ell-1} \pmod{2}$, i.e., $\bH_{\ell-1} \bc_{\ell-1}^T = \bs_{\ell-1} + 2\bb$, for some $\bb \in \ZZ^{m_{\ell-1}}$. It follows that
\begin{align*}
2^{\ell} \ba
&=2^{\ell-1} (\bH_{\ell-1} \bc_{\ell-1}^T - 2\bb) + \bH_{\ell-1}\sum_{i=0}^{\ell-2} 2^i \bc_i^\T \\
&= - 2^\ell \bb + \bH_{\ell-1}\sum_{i=0}^{\ell-1} 2^i \bc_i^\T
\end{align*}
or simply
\[
\bH_{\ell-1} \sum_{i = 0}^{\ell-1} 2^i \bc_i^\T \equiv \bzero \pmod{2^{\ell}}.
\]
Now \eqref{eq:proof-well-defined-1} follows by Lemma~\ref{lem:matrix-nesting}, completing the induction.

Since the procedure is well-defined, the set $\calCseq$ is also well-defined. We now prove that $\calCseq = \calC$. 
First, let $\bc \in \calC$. Then, for all $0\leq \ell \leq L-1$,
\begin{align*}
\bzero 
&\equiv \bH_\ell \bc^T \pmod{2^{\ell+1}} \\
&\equiv \bH_\ell \sum_{i=0}^{L-1} 2^i\bc_i^T \pmod{2^{\ell+1}} \\
&\equiv \bH_\ell 2^\ell\bc_\ell^T + \bH_\ell \sum_{i=0}^{\ell-1} 2^i\bc_i^T \pmod{2^{\ell+1}}
\end{align*}
and therefore
\begin{align*}
\bH_\ell \bc_\ell^T 
&\equiv \frac{-\bH_\ell \sum_{i=0}^{\ell-1} 2^i\bc_i^T}{2^\ell} \pmod{2} \\
&\equiv \bs_\ell \pmod{2}.
\end{align*}
Thus, $\bc$ can be generated by the procedure described, which implies $\calC \subseteq \calCseq$. Now, let $\bc \in \calCseq$. For all $0\leq \ell \leq L-1$, we have that
\[
\bH_\ell \bc_\ell^\T \equiv \bs_\ell \equiv \frac{ -\bH_{\ell} \sum_{i=0}^{\ell-1} 2^{i} \bc_{i}^\T}{2^{\ell}} \pmod{2}
\]
which implies
\[
2^\ell \bH_\ell \bc_\ell^\T + \bH_{\ell} \sum_{i=0}^{\ell-1} 2^{i} \bc_{i}^\T = 2^{\ell+1}\ba
\]
for some $\ba \in \ZZ^{m_\ell}$, and therefore
\begin{align*}
\bH_\ell \bc^T 
&\equiv \bH_\ell \sum_{i=0}^{\ell} 2^i\bc_i^T \pmod{2^{\ell+1}} \\
&\equiv \bzero \pmod{2^{\ell+1}}.
\end{align*}
It follows that $\bc \in \calC$. This proves that $\calCseq \subseteq \calC$ and thus $\calC = \calCseq$, completing the proof.
\end{IEEEproof}
\medskip

\begin{remark}
In the proof of Theorem~\ref{th:sequential_encoding}, we have only assumed the \textit{conclusion} of Lemma~\ref{lem:matrix-nesting} (property \eqref{eq:lemma1}), not the \textit{hypothesis} (definition \eqref{eq:Hmatrix}). This will be useful in Section~\ref{sec:generalized} when generalizing Construction~D$'$.
\end{remark}

\medskip

The essence of sequential encoding for Construction D$'$ is that, when encoding level $\ell$, rather than using the original linear code $\calC_\ell$, we encode using the \textit{coset code} $\calC_\ell(\bs_\ell)$ defined by the \textit{syndrome} $\bs_\ell$, which is computed based on the codewords from the previous levels. Thus, at each level, encoding can be performed entirely using a binary code.

The following result has been used in the literature (e.g., \cite{Sadeghi.etal.2006.Low-density-Parity-check-Lattices:,Choi.etal.2008.Iterative-Decoding-Low-Density}) without an explicit proof.\footnote{The classical result in \cite[Chapter 8, Theorem 14]{Conway.Sloane.1999.Sphere-Packings-Lattices} assumes that ``some rearrangement of $\bh_1,\ldots,\bh_{m_0}$ forms the rows of an upper triangular matrix,'' which allows the linear congruences to be independently solved. However, this assumption is ommitted in the definition of Construction~D' that commonly appears in the literature, such as in \cite{Sadeghi.etal.2006.Low-density-Parity-check-Lattices:,Choi.etal.2008.Iterative-Decoding-Low-Density,Kositwattanarerk.Oggier.2014.Connections-Construction-Related} and here. While it is always possible to find some $\bh_1,\ldots,\bh_{m_0}$ of this form given a family of nested codes, an explicit proof was lacking that the result still holds even for specific $\bh_1,\ldots,\bh_{m_0}$ that are not of this form.} Here, the proof follows immediately from Theorem~\ref{th:sequential_encoding}.

\medskip
\begin{corollary}
Let $\calC$ satisfy the conditions of Theorem~\ref{th:sequential_encoding}, and let $\calC_\ell$ be the null space of $\varphi(\bH_\ell)$, for $\ell = 0,\ldots,L-1$.
Then
\begin{equation}\label{eq:cardinality-total}
|\calC| = |\calC_0|\cdot \cdots \cdot |\calC_{L-1}|
\end{equation}
and therefore
\begin{equation}\label{eq:rate-total}
R(\calC) = R(\calC_0) + \cdots + R(\calC_{L-1}).
\end{equation}
\end{corollary}
\medskip

\begin{example}
\label{ex:sequential}
Let $\calC$ be the code described in Theorem~\ref{th:sequential_encoding} for $L=3$ and matrices
\begin{align*}
\bH_0 &= \mat{1 & 1 & 1 & 1 \\ 1 & 0 & 1 & 0 \\ 1 & 1 & 0 & 0} \\
\bH_1 &= \mat{1 & 1 & 1 & 1 \\ 1 & 0 & 1 & 0 } \\
\bH_2 &= \mat{1 & 1 & 1 & 1}.
\end{align*}
We construct $\bc \in \calC$ by sequential encoding through the vectors $\bc_0,\bc_1,\bc_2 \in \{0,1\}^4$. First, we choose some $\bc_0$ satisfying
\[
\bH_0 \bc_0^\T \equiv \bzero \pmod{2}
\]
for instance, $\bc_0 = (1,1,1,1)$. Then, we compute $\bs_1 = -\bH_1 \bc_0^\T / 2 \bmod 2 = (0,1)^\T$ and choose some $\bc_1$ satisfying
\[
\bH_1 \bc_1^\T \equiv \bs_1 \pmod{2}
\]
for instance, $\bc_1 = (0,1,1,0)$. Next, we compute $\bs_2 = -(\bH_2 2 \bc_1^\T + \bH_2 \bc_0^\T)/4 \bmod 2 = 0$ and choose some $\bc_2$ satisfying
\[
\bH_2 \bc_2^\T \equiv \bs_2 \pmod{2}
\]
for instance, $\bc_2 = (0,0,1,1)$. Finally, we obtain $\bc = \bc_0 + 2 \bc_1 + 4 \bc_2 = (1,1,1,1) + (0,2,2,0) + (0,0,4,4) = (1,3,7,5)$. Since $\bc = (1,3,7,5)$ satisfies all conditions in \eqref{eq:lambda_matrix_definition}, we confirm that $\bc \in \calC$.
\end{example}

\section{Encoding and Decoding}
\label{sec:codec}

In this section, we discuss conditions under which an LDPC lattice code can be encoded and decoded with constant per-bit complexity. Throughout the section, let $\calC$ be an $L$-level LDPC lattice code with component codes $\calC_\ell$ defined by parity-check matrices $\bH_\ell \in \{0,1\}^{m_\ell \times n}$, $\ell=0,\ldots,L-1$.

\subsection{Systematic Encoding}

Encoding of $\calC$ consists of bijectively mapping a tuple of message vectors
$(\bu_0,\ldots,\bu_{L-1}) \in \{0,1\}^{k_0} \times \cdots \times \{0,1\}^{k_{L-1}}$ to a codeword $\bc = \sum_{\ell=0}^{L-1} 2^{\ell} \bc_{\ell} \in \calC$. We say that the encoding is systematic if, for each $\ell$, there exists some permutation matrix $\bT_\ell$ such that, for all $\bu_\ell$, we can express $\bc_\ell = \mat{\bu_\ell & \bp_\ell}\bT_\ell$, for some $\bp_\ell \in \{0,1\}^{m_\ell}$.

Let $0\leq \ell < L$. From Theorem~\ref{th:sequential_encoding}, we know that we must have $\bc_{\ell} \in \calC_{\ell}(\bs_{\ell})$, where $\bs_\ell$ is computed from $\bc_0,\ldots,\bc_{\ell-1}$ using \eqref{eq:syndrome}. Note that $\bs_\ell$ can always be computed in $O(n)$ operations, since $\bH_\ell$ is sparse. Thus, we focus on encoding the coset code $\calC_\ell(\bs_\ell)$, for any $\bs_\ell$.

Let $\bH_\ell$ denoted by $\bH_\ell = \mat{\bH_\ell^{u} & \bH_\ell^{p}} \bT_\ell$, where $\bH_\ell^{p} \in \{0,1\}^{m_\ell \times m_\ell}$ is invertible over $\FF_2$. Note this can always be enforced by properly choosing $\bT_\ell$, since $\bH_\ell$ is assumed to be full-rank over $\FF_2$. Finding $\bc_\ell \in \calC_(\bs_\ell)$ amounts to finding $\bp_\ell \in \{0,1\}^{m_\ell}$ such that
\begin{equation*}
\bH_\ell^u \bu_\ell^\T + \bH_\ell^p \bp_\ell^\T = \bH_\ell \bc_\ell^\T \equiv \bs_\ell \pmod{2}
\end{equation*}
or, equivalently, such that
\begin{equation*}
\bH_\ell^p \bp_\ell^\T \equiv \bs_\ell - \bH_\ell^u \bu_\ell^\T \pmod{2}.
\end{equation*}
Note that $\bH_\ell^u \bu_\ell^\T$ can always be computed in $O(n)$, since $\bH_\ell$ is sparse. Thus, we have proved the following result.

\medskip
\begin{proposition}
\label{prop:efficient-encoding}
Encoding of $\calC$ can be done in $O(Ln)$ operations if the system $\bH_\ell^p \bp_\ell^\T \equiv \bs_\ell' \pmod{2}$ can be solved in $O(n)$ for all $\ell$. \end{proposition}
\medskip

One example situation where the condition of Proposition~\ref{prop:efficient-encoding} holds, as shown in \cite{Richardson.Urbanke.2001.Efficient-Encoding-Low-density}, is when each $\bH_\ell^p$ is in approximate lower triangular (ALT) form, i.e.,
\begin{equation}
\label{eq:lowertriangularform}
\bH_{\ell}^p =\mat{\bB & \bL \\ \bD & \bE}
\end{equation}
where $\bL \in \{0,1\}^{(m_\ell - g) \times (m_\ell - g)}$ is lower triangular with ones along the diagonal and $g$, called the \textit{gap} of the ALT form, is $O(1)$.

Thus, if each $\bH_\ell$ satisfies \eqref{eq:lowertriangularform} (up to row/column permutations), then the lattice code $\calC$ admits systematic encoding with complexity $O(Ln)$.

\subsection{Multistage Decoding}
\label{ssec:decoding}

Let $\bc = \sum_{\ell=0}^{L-1} 2^{\ell} \bc_{\ell} \in \calC$ be the transmitted codeword, where $\bc_{\ell}~\in~\{0,1\}^n$, and let $\br = \bc + \bz \bmod 2^L \ZZ^n$ be the received vector, where $\bz$ is a noise vector of variance $\sigma^2$ per component.

Multi-stage decoding of $\calC$ is inspired by \cite{Forney.etal.2000.Sphere-bound-achieving-Coset-Codes}. Suppose that the vectors $\bc_{\ell}$ have been correctly decoded for $i=0,1,\ldots,\mbox{$\ell-1$}$. We compute
\begin{align}
\label{eq:definition_rell}
\br_{\ell} &=  \frac{\br - \sum_{i=0}^{\ell - 1} 2^i \bc_i}{2^{\ell}}  \bmod{2} \\
\label{eq:equivalent_noise_channel}
&= \bc_{\ell} + \frac{\bz}{2^{\ell}} \bmod{2}.
\end{align}
This may be interpreted as the transmission of $\bc_\ell \in \calC_\ell(\bs_\ell)$ through a modulo-2 channel subject to additive noise $\bz/2^{\ell}$. In particular, maximum-likelihood decoding on this channel would be given by $\hat{\bc}_{\ell} = \argmax_{\bc_{\ell} \in \calC_{\ell} (\bs_{\ell})} p(\br_{\ell} | \bc_{\ell})$. It should be emphasized that re-encoding is not needed for multistage decoding, only the ability to decode a coset code. Thus, provided that each $\bs_\ell$ can be efficiently computed from the previous levels (which is always the case for LDPC lattices), efficient encoding is \textit{not} needed for efficient multistage decoding.

To obtain low complexity decoding with near optimum performance, the iterative belief propagation algorithm can be used, which has $O(n)$ complexity for sparse matrices and a limited number of iterations. The algorithm has as its input a vector $\bL\bL\bR \in \RR^n$ with the log-likelihood ratio (LLR) of each component $r_{\ell j}$ of the received vector $\br_\ell$, defined as
\begin{equation}
\label{eq:llr}
\mathrm{LLR}_j = \ln \left(\frac{ p(r_{\ell j} | c_{\ell j}=0 )}{ p(r_{\ell j} | c_{\ell j} =1 ) } \right), \quad j=1,\ldots,n.
\end{equation}

However, this algorithm assumes codewords $\bc_{\ell}$ belonging to a linear code, as opposed to an affine code $\calC_{\ell} (\bs_{\ell})$.

We can exploit this algorithm for the problem at hand using the lengthened linear code $\calC'_\ell \subseteq \FF_2^{n+m_\ell}$ defined by the parity-check matrix $\bH_{\ell}' = \mat{ -\bI & \bH_{\ell} }$, which remains sparse. In this case, the admissible codewords must be restricted to the form $\bc_{\ell}' = \mat{ \bs_\ell & \bc_{\ell} }$,  so that
\begin{equation}
\label{eq:sequential_encoding_prime}
\bH_{\ell}^{\prime} \bc_{\ell}^{\prime^{\transpose}} \equiv \bzero \pmod{2}.
\end{equation}
In order to impose this constraint, it suffices to provide as an input LLR vector the vector given as
\begin{equation}
\bL\bL\bR' = \mat{(1-2\bs_\ell) \cdot \infty & \bL\bL\bR}
\end{equation}
where the LLR value $\infty$ ($-\infty$) indicates certainty that the corresponding codeword symbol is equal to $0$ ($1$).

We conclude that the decoding of $\calC$ can be realized with complexity $O(Ln)$. By means of the union bound, the probability of error satisfies
\begin{align}
\label{eq:union_bound_multilevel}
P_e(\calC,\sigma^2) \leq {}&P_e (\calC_0,\sigma^2) + P_e \left( \calC_{1},(\sigma / 2)^2 \right) + \cdots + \nonumber \\
&+ P_e \left( \calC_{L-1},(\sigma / 2^{L-1})^2 \right)
\end{align}
where, for $0 \leq \ell \leq L-1$, $P_e (\calC_\ell,(\sigma / 2^\ell)^2)$ is the probability of error of $\calC_{\ell}$ on channel \eqref{eq:equivalent_noise_channel}.

\subsection{Multistage Decoding with Re-encoding}

Linear-time decoding can also be proved in a more general way, without relying on a specific algorithm for the coset codes, under the assumption of linear-time re-encoding of the all-zero vector.

\medskip
\begin{proposition}
Decoding of $\calC$ can be done in $O(Ln)$ operations if each linear code $\calC_\ell$ admits linear-time decoding and the condition of Proposition~\ref{prop:efficient-encoding} is satisfied.
\end{proposition}
\begin{IEEEproof}
For $\ell=0,\ldots,L-1$, let $\bv_\ell \in \calC_\ell(\bs_\ell)$ be the coset codeword corresponding to systematic encoding of the all-zero message vector $\bzero$ of length $k_\ell$ and let $\bc_\ell' \in \calC_\ell$ be the codeword corresponding to systematic encoding of the message vector $\bu_\ell \in \{0,1\}^{k_\ell}$. It follows that $\bc_\ell \in \calC_\ell(\bs_\ell)$, the coset codeword corresponding to systematic encoding of $\bu_\ell$, is given by 
\begin{equation}
\label{eq:reencoding-coset-decomposition}
\bc_\ell = \bc_\ell' + \bv_\ell \bmod 2.
\end{equation}
Now, we can modify the multistage decoding procedure to compute
\begin{align}
\br_\ell' 
&= \br_\ell - \bv_\ell \bmod 2 \\
&= \bc_\ell' + \frac{\bz}{2^\ell} \bmod 2
\end{align}
then decode $\bc_\ell' \in \calC_\ell$ and finally obtain $\bc_\ell$ with \eqref{eq:reencoding-coset-decomposition}. Note that $\bv_\ell$ can be computed by choosing $\bs_\ell' = \bs_\ell$ in Proposition~\ref{prop:efficient-encoding}. Since all the steps involved are $O(n)$, the result follows.
\end{IEEEproof}
\medskip

\section{A Generalization of Construction~D$'$}
\label{sec:generalized}

A significant limitation of Construction~D$'$ is the requirement that not only the component codes but also their corresponding parity-check matrices $\bH_\ell$ be nested, i.e., that $\bH_\ell$ be a submatrix of $\bH_{\ell-1}$. This constraint complicates the design of LDPC codes as it requires, for instance, 
that the average column weight of $\bH_{\ell-1}$ be strictly (and often significantly) higher than that of $\bH_{\ell}$, 
conflicting with the optimum design of LDPC codes for their corresponding target rates. 
In principle, one could eliminate this constraint entirely and redefine Construction~D$'$ by means of expression \eqref{eq:lambda_matrix_definition}. However, with that approach there would be no guarantee of the cardinality of $\calC$ (as given by \eqref{eq:cardinality-total}), let alone the possibility of sequential encoding, thus compromising essential properties of the construction.
The reason is that a congruence modulo $2^{\ell+1}$ in \eqref{eq:lambda_matrix_definition} applies not only to level $\ell$ but also to all levels $i < \ell$ through a reduction modulo $2^{i + 1}$. Thus, sequential encoding is not possible unless these new congruences for previous levels are completely redundant. This idea is captured by Lemma~\ref{lem:matrix-nesting}, which is the fundamental ingredient enabling sequential encoding and the guarantee of cardinality as an immediate consequence.

However, requiring that $\bH_\ell$ be a submatrix of $\bH_{\ell-1}$ is simple, but it is not the only way of satisfying Lemma~\ref{lem:matrix-nesting}. Instead, we can relax the nesting constraint on matrices $\bH_\ell$ by enforcing the more general constraint
\begin{equation}\label{eq:constraint-new}
\bH_{\ell} \equiv \bF_\ell \bH_{\ell-1} \pmod{2^{\ell}}
\end{equation}
for some integer matrix $\bF_\ell$, from which Lemma~\ref{lem:matrix-nesting} immediately follows. Clearly, requiring that $\bH_\ell$ be a submatrix of $\bH_{\ell-1}$ is a special case of this constraint.

\medskip
\begin{definition}[Generalized Construction~D$'$]
\label{def:generalized}
Let the matrices $\bH_\ell~\in~\ZZ^{m_\ell \times n}$, $\ell=0,\ldots,L-1$, be such that
\begin{enumerate}
\item $\varphi(\bH_\ell)$ is full-rank, for $\ell=0,\ldots,L-1$;
\item $\bH_{\ell}~\equiv~\bF_\ell~\bH_{\ell-1}~\pmod{2^{\ell}}$, for some  $\bF_\ell \in \ZZ^{m_\ell \times m_{\ell-1}}$, for $\ell~=~1,\ldots,L-1$.
\end{enumerate}
The lattice 
\begin{equation}\nonumber
\Lambda = \left\lbrace \bv \in \ZZ^n: \bH_\ell \bv^{\mathrm{T}} \equiv 0 \pmod{2^{\ell+1}},\; 0\leq \ell \leq L-1 \right\rbrace
\end{equation}
is said to be obtained by the Generalized Construction~D$'$ applied to $\bH_0,\ldots,\bH_{L-1}$. Equivalently, we can express $\Lambda$ as $\Lambda~=~\calC~+~2^L~\ZZ^n$, where 
$\calC = \Lambda \cap [0,2^L)^n$ is a lattice code.
\end{definition}
\medskip

It follows immediately that the set $\Lambda$ defined above is indeed a lattice.

The emphasis of Definition~\ref{def:generalized} is on the parity-check matrices $\bH_\ell$, rather than on the component codes. The interpretation based on nested component codes can be reestablished by taking $\calC_\ell~\subseteq~\{0,1\}^n$ to be such that $\varphi(\calC_\ell) \subseteq \FF_2^n$ is the null space of $\varphi(\bH_\ell) \in \FF_2^{m_\ell \times n}$. 
Clearly, as a consequence of \eqref{eq:constraint-new}, we have $\calC_0 \subseteq \calC_1 \subseteq \cdots \subseteq \calC_{L-1}$.

The main result of this section is given the by following theorem.

\medskip
\begin{theorem} Let $\calC$ be a lattice code satisfying Definition~\ref{def:generalized}. Then $\calC$ admits sequential encoding according to Theorem~\ref{th:sequential_encoding}.
Moreover, $|\calC| = |\calC_0|\cdot \cdots \cdot |\calC_{L-1}|$.
\end{theorem}
\begin{IEEEproof}
The proof follows immediately since Theorem~\ref{th:sequential_encoding} only relies on \eqref{eq:lambda_matrix_definition} and Lemma~\ref{lem:matrix-nesting}. 
\end{IEEEproof}
\medskip

It is easy to see that all of the results of Section~\ref{sec:codec} are still valid for the Generalized Construction~D$'$.

\medskip
\begin{example}\label{ex:generalized}
Let
\begin{align*}
\bF_1 &= 
\mat{
	2 & 7 & 4 \\
	11 & 9 & 6
} \\ 
\bF_2 &= \mat{3 & 5} 
\end{align*}
be arbitrarily chosen integer matrices, and let
\begin{align*}
\bH_0 &= 
\mat{
	1 & 1 & 1 & 1 \\
	1 & 0 & 1 & 0 \\
	1 & 1 & 0 & 0 
}\\
\bH_1 = \bF_1 \bH_0 \bmod{2} &= 
\mat{
	1 & 0 & 1 & 0 \\
	0 & 1 & 0 & 1
} \\
\bH_2 = \bF_2 \bH_1 \bmod{4} &= 
\mat{
	3 & 1 & 3 & 1
}.
\end{align*}
It is easy to check that $\varphi(\bH_0)$, $\varphi(\bH_1)$ and $\varphi(\bH_2)$ are full-rank.
Generalized Construction~D$'$ applied to matrices $\bH_0$, $\bH_1$, and $\bH_2$ produces a lattice code $\calC$ with $L=3$ levels and rate $R = \frac{1}{n}\log_2 |\calC|= \frac{1}{4}\log_2(2^{1 + 2 + 3}) = 1.5$ bits per dimension.

Note that $\bH_2$ is non-binary, as may be any of the matrices $\bH_\ell$, for $\ell \geq 2$. Nevertheless, all the encoding and decoding operations are still performed over $\FF_2$ with $\varphi(\bH_\ell)$, except for the computation of the syndrome $\bs_{\ell}$ in \eqref{eq:syndrome}.
\end{example}

\medskip
The following theorem shows that the only real requirement for Generalized Construction~D$'$ is that the linear component codes be nested. Any (full-rank) parity-check matrices for these codes, nested or not, may be used in the construction, although we may need to lift them to non-binary integers.

\medskip
\begin{theorem}
Let $\calC_0 \subseteq \calC_1 \subseteq \cdots \subseteq \calC_{L-1} \subseteq \{0,1\}^n$ be nested linear codes with parity-check matrices $\bar{\bH}_\ell \in \{0,1\}^{m_\ell \times n}$, $\ell = 0,\ldots,L-1$, respectively. Then there exist matrices $\bH_\ell \in \ZZ^{m_\ell \times n}$, $\ell=0,\ldots,L-1$, satisfying \eqref{eq:constraint-new} and such that $\bH_\ell \equiv \bar{\bH}_\ell \pmod{2}$.
\end{theorem}
\begin{IEEEproof}
Let $\bH_0 = \bar{\bH}_0$. For $\ell=1,\ldots,L-1$, we proceed by induction. Assume that $\bH_{\ell-1} \equiv \bar{\bH}_{\ell-1} \bmod 2$, which is true for $\ell=1$. Since $\calC_{\ell-1} \subseteq \calC_{\ell}$, we have that $\calC_{\ell}^\perp \subseteq \calC_{\ell-1}^\perp$, where $\calC_\ell^\perp$ denotes the dual code of $\calC_\ell$. This implies that there exists some $\bF_\ell \in \{0,1\}^{m_\ell \times m_{\ell-1}}$ such that $\bar{\bH}_\ell \equiv \bF_\ell \bar{\bH}_{\ell-1} \bmod 2$. Let $\bH_\ell = \bF_\ell \bH_{\ell-1} \bmod 2^\ell$, which automatically satisfies \eqref{eq:constraint-new}. It follows that
\begin{align}
\bH_\ell 
&\equiv \bF_\ell \bH_{\ell-1} \bmod 2 \\
&\equiv \bF_\ell \bar{\bH}_{\ell-1} \bmod 2 \\
&\equiv \bar{\bH}_{\ell} \bmod 2
\end{align}
completing the induction.
\end{IEEEproof}

\subsection{Comparison with Construction~D$'$}

In the remainder of this section, we compare Construction~D$'$ and Generalized Construction~D$'$ under two common perspectives, 
which differ essentially on whether complexity is taken into account.

\subsubsection{As a Codebook Construction}

From a purely theoretical (or geometric) perspective, a code or lattice is defined as a set of points in some space, i.e., as a codebook. In this case, one is concerned with geometric properties of the codebook such as minimum distance or probability of error under minimum distance decoding. This is the approach implicit in classical descriptions of linear codes and lattices \cite{Conway.Sloane.1999.Sphere-Packings-Lattices} and, for instance, in the comparison between Constructions D and D$'$ in \cite{Kositwattanarerk.Oggier.2014.Connections-Construction-Related}.

From this perspective, Generalized Construction~D$'$ is strictly more general than Construction~D$'$, as there exist examples of the former that cannot be described by the latter. Considering an $L$-level Generalized Construction~D$'$ lattice with matrices $\bH_0,\ldots,\bH_{L-1}$, such examples can always be produced if $L>2$ or if $\bH_1 \bmod 4$ is \textit{non-binary}. Otherwise, if $L=2$ and $\bH_1 \bmod 4$ is binary, then the same codebook can be produced using Construction~D$'$ with matrices $\bar{\bH}_0$ and $\bar{\bH}_1$, where $\bar{\bH}_1 = \bH_1 \bmod 2$ and $\bar{\bH_0}$ is any binary matrix that defines $\calC_0$ and contains $\bH_1$ as a submatrix. This follows since, as can be seen from Theorem~\ref{th:sequential_encoding},  both constructions depend only on $\calC_0$ and $\bH_\ell \bmod 2^{\ell+1}$, $\ell=1,\ldots,L-1$. 

Examples of the non-equivalent cases are shown next.

\medskip
\begin{example}
Consider again Examples~\ref{ex:sequential} and~\ref{ex:generalized}, but let all matrices, syndromes, codewords and lattice code of Example~\ref{ex:sequential} be denoted with an overline, such as $\bar{\bH}_\ell$, $\bar{\bs}_\ell$, $\bar{\bc}_\ell$ and $\bar{\calC}$, to distinguish them from those of Example~\ref{ex:generalized}.
Clearly, the underlying binary codes are the same, namely,
\begin{align*}
\calC_0 &= \langle(1,1,1,1)\rangle \\
\calC_1 &= \langle(1,1,1,1), (1,0,1,0)\rangle \\
\calC_2 &= \langle(1,1,1,1), (1,0,1,0), (0,0,1,1) \rangle.
\end{align*}
However, in contrast to $\bar{\bH}_0$, $\bar{\bH}_1$ and $\bar{\bH}_2$, neither $\bH_1$ is a submatrix of $\bH_0$, nor $\bH_2$ is a submatrix of $\bH_1$.
We will construct a codeword $\bc = \bc_0 + 2\bc_1 + 4\bc_2 \in \calC$ such that $\bc \not\in \bar{\calC}$.

Let $\bc_0 = \bar{\bc}_0 = (1,1,1,1)$. Then $\bs_1 = (1,1)^T$, from which we may choose $\bc_1 = (1,1,0,0) \in \calC_1(\bs_1)$. However, as noted before, $\bar{\bs}_1 = (0,1)^T \neq \bs_1$, which implies that 
$\bar{\bc}_1$ must be chosen from a different coset of $\calC_1$ and therefore it cannot equal $\bc_1$.
Thus, necessarily $\bc \not\in \bar{\calC}$, regardless of $\bc_2$.

Since $\bH_1$ is binary, one may attempt to work around this problem by redefining $\bar{\bH}_1  = \bH_1$ and \begin{equation*}
\bar{\bH}_0 = \mat{
	1 & 0 & 1 & 0 \\
	0 & 1 & 0 & 1 \\
	1 & 1 & 0 & 0
}
\end{equation*}
so that $\bar{\bH}_1$ is still a submatrix of $\bar{\bH}_0$, but now $\bar{\bs}_1 = \bs_1$. Thus, we can choose $\bar{\bc}_1 = \bc_1$. However, $\bar{\bH}_2$ must be chosen as a submatrix of $\bar{\bH}_1$ and no such choice can produce $\calC_2$; for instance, the vector $(0,0,1,1) \in \calC_2$ will never be in the null space of $\varphi(\bar{\bH}_2)$.

More generally, 
the code $\calC_2$ defined by $\bH_2$
can only be produced with $\bar{\bH}_2 = \mat{1 & 1 & 1 & 1}$, which in turn implies that $(1,1,1,1)$ must be a row of $\bar{\bH}_1$. Hence, $\bar{\bc}_0$ must produce a syndrome $\bar{\bs}_1$ with at least one zero entry, and thus $\bar{\bs}_1 \neq \bs_1$. It follows that necessarily $\bar{\calC} \neq \calC$, i.e., the lattice code $\calC$ cannot be produced by Construction~D$'$.
\end{example}
\medskip

\begin{example} 
Let $L=2$ and
\begin{align*}
\bF_1
&= \mat{3 & 1} \\
\bH_0
&= \mat{1 & 0 & 0 & 1 \\
		1 & 1 & 0 & 0
		} \\
\bH_1 
= \bF_1 \bH_0 \bmod{4}
&= \mat{0 & 1 & 0 & 3}.
\end{align*}
Clearly, we still have $\bH_1 \equiv \bF_1 \bH_0 \pmod{2}$. Let $\calC \subseteq [0,4)^4$ be the lattice code produced by Generalized Construction~D$'$ with matrices $\bH_0$ and $\bH_1$.
The underlying nested codes are
\begin{align*}
\calC_0 &= \langle(0,0,1,0), (1,1,0,1)\rangle \\
\calC_1 &= \langle(0,0,1,0), (1,1,0,1), (1,0,0,0)\rangle.
\end{align*}
Let $\bar{\bH}_0 \in \{0,1\}^{2 \times 4}$ and $\bar{\bH}_1 \in \{0,1\}^{1 \times 4}$ be nested matrices that define the same codes $\calC_0$ and $\calC_1$, respectively, and let $\bar{\calC} \subseteq [0,4)^4$ be the corresponding lattice code produced by Construction~D$'$ with $\bar{\bH}_0$ and $\bar{\bH}_1$. Clearly, there is a single possibility for $\bar{\bH}_1$, namely, 
\begin{equation*}
\bar{\bH}_1 = \mat{0 & 1 & 0 & 1}.
\end{equation*}
Let $\bc_0 = (1,1,1,1)$. Then the corresponding syndrome at level $1$, when computed with $\bH_1$, is equal to $\bs_1 = 0$ but, when computed with $\bar{\bH}_1$, it is equal to $\bar{\bs}_1 = 1$. Thus, for any valid choice of $\bar{\calC}$, we have $\bc=(1,1,1,1) \in \calC$ but $\bc \not\in \bar{\calC}$.
\end{example}
\medskip

It is worth pointing out that the greater flexibility of Generalized Construction~D$'$ does not offer any advantage in terms of \textit{theoretical} performance (with unbounded complexity) under multistage decoding. This is because the error probability in \eqref{eq:union_bound_multilevel} depends solely on the component codes $\calC_0,\ldots,\calC_{L-1}$, and we can always produce a Construction~D$'$ lattice with the same component codes. In this case, as shown in the examples above, the only difference between the two constructions is in the syndrome calculation at each level.

More generally, any multilevel codes that share the same component codes (including Construction~D lattices) will also have the same error probability under multistage decoding if we 
allow unbounded complexity.

\subsubsection{As a Construction of a Coding Scheme}

From a practical (or complexity-constrained) perspective, a coding scheme consists of a codebook, an encoding function and a decoding function, and one is concerned, in particular, with the performance of the scheme under a certain complexity. The decoder thus plays a key role in the construction of the scheme. This is the approach implicit in descriptions of modern codes such as Turbo and LDPC codes \cite{Richardson.Urbanke.2008.Modern-Coding-Theory}; in particular, 
an LDPC code is described not simply as a set of codewords, but through some specific parity-check matrix that induces a convenient decoder structure.

This second perspective is the focus of this paper and is what motivates our definition of Generalized Construction~D$'$. From this perspective, assuming that the decoder is specified by the parity-check matrices used in the lattice construction, Generalized Construction~D$'$ is indeed more general than Construction~D$'$, since it introduces fewer constraints on the choice of the parity-check matrices. This increased flexibility can translate into a better performance if the decoder is sensitive to the choice of the parity-check matrices, which is the case of LDPC lattices under a multistage belief-propagation decoder. Numerical examples of their difference in performance are shown in Section~\ref{ssec:results-fer5-n1024}.

One way to artificially match the performance of the two constructions may be to first design a Generalized Construction~D$'$ lattice with matrices $\bH_0,\ldots,\bH_{L-1}$ and then create a Construction~D$'$ lattice with nested matrices $\bar{\bH}_0,\ldots,\bar{\bH}_{L-1}$ that correspond to the same component codes. Then, use $\bar{\bH}_0,\ldots,\bar{\bH}_{L-1}$ for encoding 
and demapping from a codeword to a message vector, while using $\bH_0,\ldots,\bH_{L-1}$ solely for ``denoising,'' i.e., for decoding from the received vector to a codeword. A clear disadvantage of this 
approach is that, for each component code, two distinct parity-check matrices must be stored and used, while, with Generalized Construction~D$'$, a single matrix can be used in all encoding and decoding steps. Moreover, it is unclear whether the nested matrices $\bar{\bH}_0,\ldots,\bar{\bH}_{L-1}$ (of which we have less control) will have a convenient structure for efficient encoding and demapping.
In other words, requiring the parity-check matrices to be nested is an unnecessary constraint of Construction~D$'$, both in theory and in practice.

More fundamentally, Construction~D$'$ was originally defined \cite{Barnes.Sloane.1983.New-Lattice-Packings,Conway.Sloane.1999.Sphere-Packings-Lattices} with a focus on minimum distance, i.e., on packing density, regardless of the availability of an efficient decoder, so allowing for a flexible choice of parity-check matrices was unnecessary. In contrast, the approach of Generalized Construction~D$'$ allows the decoder structure, embodied by specific parity-check matrices, to be taken into account as part of the design.

\section{Nested LDPC Codes by Check Splitting}
\label{sec:split}

In this section, we propose a method to construct suitable \textit{binary} matrices $\bH_0,\ldots,\bH_{L-1}$ that satisfy the conditions of Generalized Construction D$'$. Our approach is to sequentially construct matrix $\bH_{\ell-1}$ based on matrix $\bH_{\ell}$, for $\ell = \mbox{$L-1$},\ldots,1$, assuming we are given the parity-check matrix $\bH_{L-1}$ of the highest-rate code and the desired number of rows for the remaining matrices, $m_{L-2},\ldots,m_{0}$.

Our method guarantees that, for each new level, the column weights of the initial matrix are preserved, i.e., all component codes will share the same variable-node degree distribution. While this approach may not lead to an optimal design for all rates, it allows us to choose, for instance, all component codes to be variable-regular LDPC codes with variable-node degree $d_v = 3$, which we can expect to exhibit at least a reasonable performance. This choice is simply not available with the original Construction~D$'$.

We also present variations of the basic method aimed at maximizing girth and at allowing linear-time encoding, inspired by the progressive edge growth (PEG) algorithm \cite{Hu.etal.2005.Regular-Irregular-Progressive}.

For the remainder of the section, it suffices to consider the following problem: given a full-rank matrix $\bB \in \{0,1\}^{b \times n}$ and a desired number of rows $m > b$, construct a full-rank matrix $\bH \in \{0,1\}^{m \times n}$ with the same sequence of column weights and such that $\bB = \bF \bH$, for some $\bF \in \ZZ^{b \times m}$. Note that we require equality over $\ZZ$, which immediately implies \eqref{eq:constraint-new} for any $\ell$.

Since an equivalent representation of a binary matrix $\bH$ is a Tanner graph (a bipartite graph, with $m$ check nodes and $n$ variable nodes, having $\bH \in \{0,1\}^{m \times n}$ as incidence matrix), we use the two concepts interchangeably.

\subsection{Check Splitting}

\begin{figure}
	\begin{center}
		\includegraphics[height=7cm]{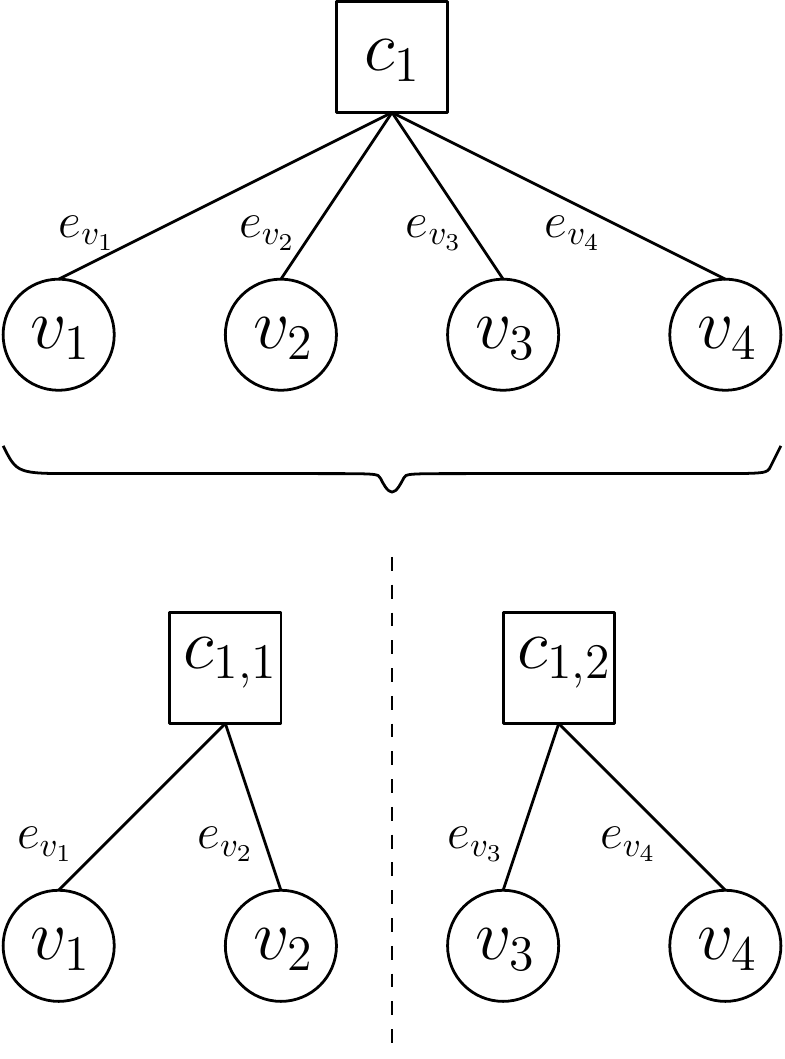}
	\end{center}
	\caption{Example of check splitting. Check node $c_1$ is split into two check nodes $c_{1,1}$ and $c_{1,2}$. The number of edges is preserved and each edge remains incident on the same variable node.}
		\label{fig:check_split}
\end{figure}

We first describe a general method based on the splitting of parity-check equations, which we refer to as \textit{check splitting}\footnote{Our definition of check splitting differs from that in \cite{Good.Kschischang.2006.Incremental-Redundancy-Check}, which introduces a variable node connecting the check nodes produced from splitting.} for short. The basic idea is illustrated in Fig.~\ref{fig:check_split}, where a check node is split in two without changing the variable nodes on which the corresponding edges are incident.

For all $\bH = [\bH(i,j)] \in \{0,1\}^{m \times n}$, let $\calJ(\bH) = (\calJ_1(\bH), \ldots, \calJ_m(\bH))$, where each $\calJ_i(\bH) = \{j \in \{1,\ldots,n\}: \bH(i,j) \neq 0\}$ is a set containing the indices of the nonzero entries of the $i$th row of~$\bH$. Similarly, let $\calI(\bH) = \left( \calI_1(\bH), \ldots, \calI_n(\bH) \right)$, where each $\calI_j(\bH) = \{i \in \{1,\ldots,m\}: \bH(i,j) \neq 0 \}$ is a set containing the indices of the nonzero entries of the $j$th column of $\bH$. 

Let $m \geq b$ and let $p: \{1,\ldots,m\} \to \{1,\ldots,b\}$ be a surjective mapping. A matrix $\bH \in \{0,1\}^{m \times n}$ is said to be obtained from $\bB \in \{0,1\}^{b \times n}$ by check splitting based on the parent mapping $p$ if, for all $i=1,\ldots,b$, the set $\{\calJ_i(\bH): i \in p^{-1}(k)\}$ forms a partition of $\calJ_k(\bB)$, where $p^{-1}(k) = \{i \in \{1,\ldots,m\}: p(i)=k\}$ is the preimage of $k$ under $p$.

Clearly, check splitting preserves column weights, since every nonzero entry of $\bB$ appears in $\bH$ in the same column, although possibly in a different row.
Moreover, it is easy to see that $\bB = \bF \bH$, where $\bF \in \{0,1\}^{b \times m}$ is such that $\calJ(\bF) = (p^{-1}(1),\ldots,p^{-1}(b))$, i.e., adding the rows of $\bH$ with indices in $p^{-1}(k)$ gives precisely the $k$th row of $\bB$.

\medskip
\begin{example}\label{ex:split}
Starting with
\[
\bH_2 = \mat{1 & 1 & 1 & 1 & 1 & 1 & 1 & 1}
\]
we can partition it into
\[
\bH_1 = \mat{
1 & 0 & 0 & 1 & 0 & 1 & 1 & 0 \\
0 & 1 & 1 & 0 & 1 & 0 & 0 & 1 
}
\]
which, in turn, can be partitioned into 
\[ \bH_0 = 
\mat{
	0 & 0 & 0 & 1 & 0 & 1 & 0 & 0 \\
	1 & 0 & 0 & 0 & 0 & 0 & 1 & 0 \\
	0 & 1 & 0 & 0 & 0 & 0 & 0 & 1 \\
	0 & 0 & 1 & 0 & 1 & 0 & 0 & 0
}.
\]
It is easy to check that $\bH_1 = \bF_1 \bH_0$ and $\bH_2 = \bF_2 \bH_1$, where
\begin{align*}
\bF_1 &= \mat{
1 & 1 & 0 & 0 \\
0 & 0 & 1 & 1
} \\
\bF_2 &= \mat{1 & 1}.
\end{align*}
Note that all matrices are binary and that each column has the same weight, namely $1$.
\end{example}
\medskip

A useful property of check splitting, which follows from \cite[Lemma 8]{Subramanian.etal.2011.Strong-Secrecy-Binary}, is that it cannot reduce girth. Thus, in particular, $\bH$ is guaranteed to be free of 4-cycles if $\bB$ is so.

\subsection{PEG-Based Check Splitting}

While girth preservation is a desirable feature, one typically expects a lower-rate code to have a better cycle distribution than a higher-rate code of the same length, preferably a larger girth. This is possible in the check splitting procedure if checks are split in a way that breaks short cycles in which they are involved. More generally, it is conceivable that a judicious choice of check splits may increase the performance of the resulting code.

In the following, we propose a check splitting algorithm, inspired by the PEG algorithm \cite{Hu.etal.2005.Regular-Irregular-Progressive}, that attempts to maximize the girth of the resulting matrix. For all $i=1,\ldots,m$ and all $j=1,\ldots,n$, let $d_\bH(i,j)$ denote the distance from check node~$i$ to variable node~$j$ in the Tanner graph induced by $\bH$, where $d_\bH(i,j) = \infty$ if there is no path joining these nodes. As shown in Algorithm~\ref{alg:peg_check_splitting}, the proposed method greedly processes each $(k,j)$ nonzero entry of $\bB$, adding a corresponding entry in $\bH$ in the same column $j$ but in some child row $i \in p^{-1}(k)$ that satisfy two goals: first, it should maximize the distance to variable node $j$ of the resulting Tanner graph; second, if multiple possibilities remain, a check of lowest degree is chosen. The algorithm can be interpreted as a generalization of the PEG algorithm where, for each iteration, the set of allowed checks is restricted to $p^{-1}(k)$, as shown in line~\ref{line:children-checks}. (The original PEG algorithm is essentially recovered if this line is replaced by $\calI \leftarrow \{1,\ldots,m\}$.)
\begin{algorithm}[t]
\caption{PEG-based check splitting}
\label{alg:peg_check_splitting}
\begin{algorithmic}[1]
\REQUIRE $\bB \in \{0,1\}^{b \times n}$, $m$
\ENSURE $\bH \in \{0,1\}^{m \times n}$
\STATE Create the parent mapping $p$.
\STATE Initialize $\bH \leftarrow \bzero$.
\FOR{$j = 1, \ldots, n$}
\FOR{$k \in \calI_j(\bB)$}
\STATE $\calI \leftarrow p^{-1}(k)$	\label{line:children-checks}
\STATE $\calI \leftarrow \{i \in \calI: d_\bH(i,j) = \max_{i' \in \calI} d_\bH(i',j)\}$
\STATE $\calI \leftarrow \{i \in \calI: |\calJ_i(\bH)| = \min_{i' \in \calI} |\calJ_{i'}(\bH)|\}$
\STATE Choose some $i \in \calI$ and set $\bH(i,j) \leftarrow 1$.
\ENDFOR
\ENDFOR
\end{algorithmic}
\end{algorithm}

A secondary goal of the algorithm is to make the weights of the resulting rows as uniform as possible, which is also a desirable feature of a good LDPC code. However, some care must be taken to ensure that the parent mapping $p$ is indeed suitable to result in concentrated weights, which is accomplished by Algorithm~\ref{alg:create_parent_mapping}. Specifically, for each $i$th row of $\bH$, this algorithm chooses as its parent row in $\bB$ one that maximizes the metric
\begin{equation}
\mu_{p}(k) \triangleq \frac{|\calJ_k(\bB)|}{|p^{-1}(k)|+1}
\end{equation}
which can be interpreted as the average weight of a corresponding child row after the parent assignment is made (i.e., 
a row is chosen that maximizes the resulting weight after a further split).
\begin{algorithm}[t]
\caption{Create the parent mapping for Algorithm~\ref{alg:peg_check_splitting}}
\label{alg:create_parent_mapping}
\begin{algorithmic}[1]
\REQUIRE $\bB \in \{0,1\}^{b \times n}$, $m$
\ENSURE $p: \{1,\ldots,m\} \to \{1,\ldots,b\}$
\STATE Set $p(i) = i$ for $i = 1,\ldots,b$.
\FOR{$i=b+1,\ldots,m$}
\STATE $\calK \leftarrow \{1,\ldots,m\}$ \label{line:parent-mapping-full}
\STATE $\calK \leftarrow \{k \in \calK: \mu_p(k) = \max_{k' \in \calK} \mu_p(k')\}$
\STATE Choose some $k \in \calK$ and set $p(i) \leftarrow k$.
\ENDFOR
\end{algorithmic}
\end{algorithm}

\subsection{Triangular PEG-Based Check Splitting}
\label{ssec:triangular-peg-cs}

A drawback of Algorithm~1 is that it generally does not produce matrices that have an approximate triangular structure. We propose a simple adaptation that ensures such a structure, thereby enabling efficient encoding. The algorithm takes a base matrix $\bB$ 
in ALT form
with gap $g$ and returns a check-split matrix $\bH$ in the same form and with the same gap $g$. Thus, $\bB$ and $\bH$ will have exactly the same encoding complexity.

For ease of notation, we assume that $\bB$ is actually in approximate \textit{upper} triangular form, which can be easily accomplished by left-right and up-down flipping of a matrix in 
ALT form.
The resulting matrix $\bH$ is similarly given in the same form.

The proposed method, which is again inspired from \cite{Hu.etal.2005.Regular-Irregular-Progressive}, is given in Algorithm~\ref{alg:triangular_check_splitting}. The differences from Algorithm~\ref{alg:peg_check_splitting} are essentially the inclusion of lines~\ref{line:add-diagonal-edge-start}--\ref{line:add-diagonal-edge-end}, which add a $1$ in the $g$th subdiagonal, and the modification in line~\ref{line:children-checks-triangular}, which restrict the valid choices of child rows to those above the $g$th subdiagonal. The crucial assumption is that $p(g+j) \in \calI_j(\bB)$ for all $j=1,\ldots,m-g$, since a $1$ can only be added in position $(g+j,j)$ of $\bH$ if a $1$ exists in position $(k_0,j)$ of $\bB$, where $k_0 = p(g+j)$ is the corresponding parent row. This property can be ensured during the creation of the parent mapping by a simple modification in line~\ref{line:parent-mapping-full} of Algorithm~\ref{alg:create_parent_mapping}, as shown in Algorithm~\ref{alg:create_parent_mapping_triangular}.

\begin{algorithm}[t]
\caption{Triangular PEG-based check splitting}
\label{alg:triangular_check_splitting}
\begin{algorithmic}[1]
\REQUIRE $\bB \in \{0,1\}^{b \times n}$, $g$, $m$
\ENSURE $\bH \in \{0,1\}^{m \times n}$
\STATE Create the parent mapping $p$.
\STATE Initialize $\bH \leftarrow \bzero$.
\FOR{$j = 1, \ldots, n$}
\STATE $\calK \leftarrow \calI_j(\bB)$
\IF{$j \leq m-g$} \label{line:add-diagonal-edge-start}
    \STATE $\bH(g+j,j) \leftarrow 1$
\STATE $k_0 \leftarrow p(g+j)$
\STATE $\calK \leftarrow \calK \setminus \{k_0\}$
\ENDIF \label{line:add-diagonal-edge-end}
\FOR{$k \in \calK$}
\STATE $\calI \leftarrow p^{-1}(k) \cap \{1,\ldots,\min\{g+j-1,m\}\}$ \label{line:children-checks-triangular}	
\STATE $\calI \leftarrow \{i \in \calI: d_\bH(i,j) = \max_{i' \in \calI} d_\bH(i',j)\}$
\STATE $\calI \leftarrow \{i \in \calI: |\calJ_i(\bH)| = \min_{i' \in \calI} |\calJ_{i'}(\bH)|\}$
\STATE Choose some $i \in \calI$ and set $\bH(i,j) \leftarrow 1$.
\ENDFOR
\ENDFOR
\end{algorithmic}
\end{algorithm}

\begin{algorithm}[t]
\caption{Create the parent mapping for Algorithm~\ref{alg:triangular_check_splitting}}
\label{alg:create_parent_mapping_triangular}
\begin{algorithmic}[1]
\setcounter{ALC@line}{2}
\STATE $\calK \leftarrow \calI_{i-g}(\bB)$
\end{algorithmic}
(Note: only the differences from Algorithm~\ref{alg:create_parent_mapping} are shown.)
\end{algorithm}

\section{Simulation Results}
\label{sec:results}

This section discusses the design and error performance simulation of LDPC lattices with \mbox{$L=2$} coded levels.

In all scenarios considered, we construct LDPC lattices $\Lambda = \calC + 4\ZZ^n$ with the Generalized Construction~D$'$ applied to matrices $\bH_0 \in \{0,1\}^{m_0 \times n}$, and $\bH_1 \in \{0,1\}^{m_1 \times n}$, corresponding to the nested codes $\calC_0~\subseteq~\calC_1$ of rates $R_0$ and $R_1$, respectively. Both codes are chosen to be variable-regular LDPC codes with variable-node degree $d_v = 3$. Matrix $\bH_1$ is constructed via the triangular version of the PEG algorithm \cite{Hu.etal.2005.Regular-Irregular-Progressive}, modified to allow for a gap $g$, which is chosen to be $g=22$, while $\bH_0$ is obtained from $\bH_1$ via triangular PEG-based check splitting with the same gap. Thus, $\calC$ is guaranteed to have efficient encoding.

Transmission over an unconstrained AWGN channel with noise variance $\sigma^2$ is simulated using the approach of Section~\ref{ssec:forney}. Decoding of the lattice code $\calC$ is performed as described in Section~\ref{ssec:decoding}, where for each component code the belief propagation decoder performs a maximum of 50 iterations. Each simulation point is obtained after the occurrence of at least $100$ word errors, except for points with word error rate (WER) below $10^{-6}$, which were obtained with at least $50$ word errors.

Decoding of the sublattice $\Lambda' = 4\ZZ^n$ is not simulated; instead, $P_e(4\ZZ^n,\sigma^2)$ is computed analytically from \eqref{eq:an-Pe-4ZZ} and applied to \eqref{eq:union_bound}, which is assumed to hold with equality. Note that, by combining \eqref{eq:union_bound} and \eqref{eq:union_bound_multilevel}, we have 
\begin{align}
\label{eq:union_bound_unconstrained}
P_e (\Lambda, \sigma^2) \leq &P_e(\calC_0,\sigma^2) +  P_e(\calC_1,\left(\sigma/2\right)^2) + P_e(4\ZZ^n,\sigma^2).
\end{align}

\subsection{Design for $P_e \leq 10^{-5}$ and $n=1024$}
\label{ssec:results-fer5-n1024}

For our first example, we have used the same design parameters as the 2-level polar lattice of \cite{Yan.etal.2014.Construction-Capacity-achieving-Lattice}, namely $P_e \leq 10^{-5}$ and $n=1024$.
For the choice of $m_0$ and $m_1$, or, equivalently, $R_0$ and $R_1$, we have not made any attempt at optimizing for a target error probability, but simply adopted the same values obtained in \cite{Yan.etal.2014.Construction-Capacity-achieving-Lattice}, namely $R_0~=~0.23$ ($m_0 = 788$) and $R_1~=~0.90$ ($m_1 = 103$), in order to allow for a simpler comparison. 

The rate design in \cite{Yan.etal.2014.Construction-Capacity-achieving-Lattice} was done using the equal error probability rule \cite{Wachsmann.etal.1999.Multilevel-Codes:-Theoretical} applied to the union-bound estimate of equation \eqref{eq:union_bound_unconstrained}. Under this rule, the goal is to make the error probability of the three levels equal (in this case, equal to $10^{-5}/3$). Using \eqref{eq:an-Pe-4ZZ}, uncoded level 2 achieves $P_e = 10^{-5}/3$ at $\sigma = 0.3380$, yielding a design $\vnr$ of $2.34$~dB. 

\begin{figure}
	\begin{center}
		\includegraphics[width=0.5\textwidth]{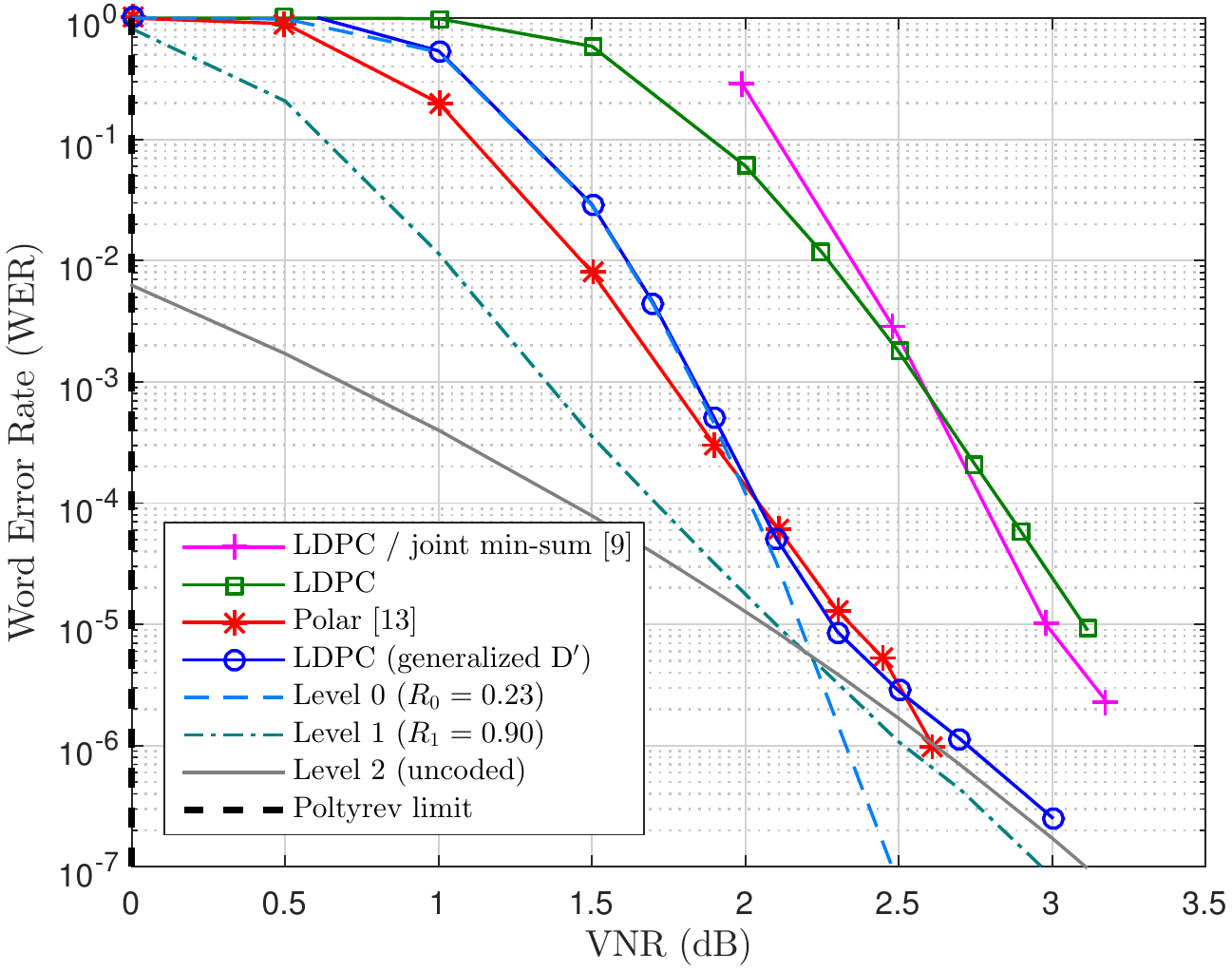}
	\end{center}
	\caption{Performance of 2-level LDPC lattices of dimension $1024$ via Generalized Construction~D$'$ and via the original Construction~D$'$, designed to achieve $P_e \leq 10^{-5}$ under multistage decoding. For comparison, the performance of a 2-level polar lattice with $n=1024$ \cite{Yan.etal.2014.Construction-Capacity-achieving-Lattice}, a 2-level LDPC lattice with $n=1000$ decoded with joint min-sum \cite{Sadeghi.etal.2006.Low-density-Parity-check-Lattices:}, and the Poltyrev limit are also shown.}
	\label{fig:unconstrained_1e-5}
\end{figure}

Fig. \ref{fig:unconstrained_1e-5} 
shows the word error rate as a function of the $\vnr$ for the Generalized Construction~D$'$ LDPC lattice and for the polar lattice. As can be seen, at the design VNR, the performance of both lattices is comparable. In particular, the LDPC lattice attains $\wer = 10^{-5}$ at  $\vnr = 2.2865$~dB.

In order to better understand the performance of the LDPC lattice, Fig. \ref{fig:unconstrained_1e-5} also shows the performance of each individual level computed without error propagation. As we can see, the fact that the slope of the curve becomes less steep after $\wer=10^{-5}$ is due to the performance of level 1 and especially of level 2. Thus, the error rate of level 2 induces a lower bound on the performance of the LDPC lattice. Note that the polar lattice has the exact same uncoded level and the same fundamental volume, being therefore limited by the same lower bound. As illustrated in the next subsection, this dependence on the uncoded level can be mitigated by an improved rate design procedure.

To illustrate the motivation for the Generalized Construction~D$'$, Fig. \ref{fig:unconstrained_1e-5} shows the performance of a $(2,3;4)$-regular LDPC lattice with $n=1000$ from \cite{Sadeghi.etal.2006.Low-density-Parity-check-Lattices:}, where all levels are decoded jointly using the min-sum algorithm. Fig. \ref{fig:unconstrained_1e-5} also shows the performance of our best attempt at designing an LDPC lattice via the original Construction~D$'$, where $\bH_0$ is constrained to be a submatrix of $\bH_1$, but instead using a low-complexity multistage decoder. In this case, both matrices were constructed together via the extended PEG algorithm from \cite{Sadeghi.etal.2006.Low-density-Parity-check-Lattices:}. We used the same design criterion as before (equal error probability under multistage decoding). Assuming variable-regular degree distributions, our best design was found with degrees $d_v^0 = 6$ and $d_v^1 = 3$ and rates $R_0 = 0.0967$ ($m_0 = 925$) and $R_1 = 0.9043$ ($m_1 = 98$). Compared to our previous design, the design of $\calC_1$ remained almost unchanged, while $R_0$ had to be significantly decreased in order for $\calC_0$ to meet the desired error rate. This poor performance of $\calC_0$ may be explained by its highly suboptimal degree distribution constrained by $\bH_1$.

As we can see, the approach of Section~\ref{sec:codec} allows us to a obtain a performance similar to that of \cite{Sadeghi.etal.2006.Low-density-Parity-check-Lattices:}, but with a lower decoding complexity. On the other hand, both lattices display a significant performance gap compared to polar lattices, as well as to the LDPC lattice constructed with Generalized Construction~D$'$.

\subsection{Design for $P_e \leq 10^{-2}$ and $n=1000$}
\label{ssec:1e-2_n1000}

For our second design example, we use $P_e \leq 10^{-2}$ and $n=1000$ as design parameters. However, we now adopt an optimized rate design procedure.

Let $\calC (R) \subseteq \{0,1\}^n$ denote a family of LDPC codes parameterized by their rate $R$. Define
\begin{equation*}
f(R,\sigma) \triangleq P_e (\calC (R), \sigma^2 ).
\end{equation*} 
as a function of the rate $R$ and noise level $\sigma$.
Our proposed design rule selects the code rates that minimize the VNR such that the error probability is kept less than or equal to $P_e$.

Using the fact that 
\begin{equation*}
\vnr = \frac{ 2^{n \left( L-\sum_{\ell=0}^{L-1} R_\ell \right) } }{ 2\pi e \sigma^2 }
\end{equation*}
this optimization problem can be rewritten as
\begin{align}
\label{eq:desgin_rule_ML}
&\{R_0^*,R_1^*,\sigma^*\} = \argmax\limits_{R_0,R_1,\sigma^2} \quad R_0 + R_1 + \log_2{\sigma} \\[1ex]
&\textrm{s.t.} \quad f(R_0,\sigma) +  f(R_1,\sigma/2) + P_e (4\ZZ^n,\sigma^2) \leq P_e. \nonumber 
\end{align}

The function $f(R,\sigma)$ is computed numerically by constructing a code $\calC(R)$ and estimating its error probability at noise level $\sigma$ via simulation. To alleviate the complexity of this estimation, simulation is used only for certain values of $R$ and $\sigma$ and linear regression is used to interpolate between any other values required by the optimization algorithm.

With this design rule, rates $R_0 = 0.5$ ($m_0 = 500$) and $R_1~=~0.978$ ($m_1 = 22$) are obtained, yielding $\vnr = 1.356$~dB for $P_e = 10^{-2}$.

\begin{figure}
	\begin{center}
		\includegraphics[width=0.5\textwidth]{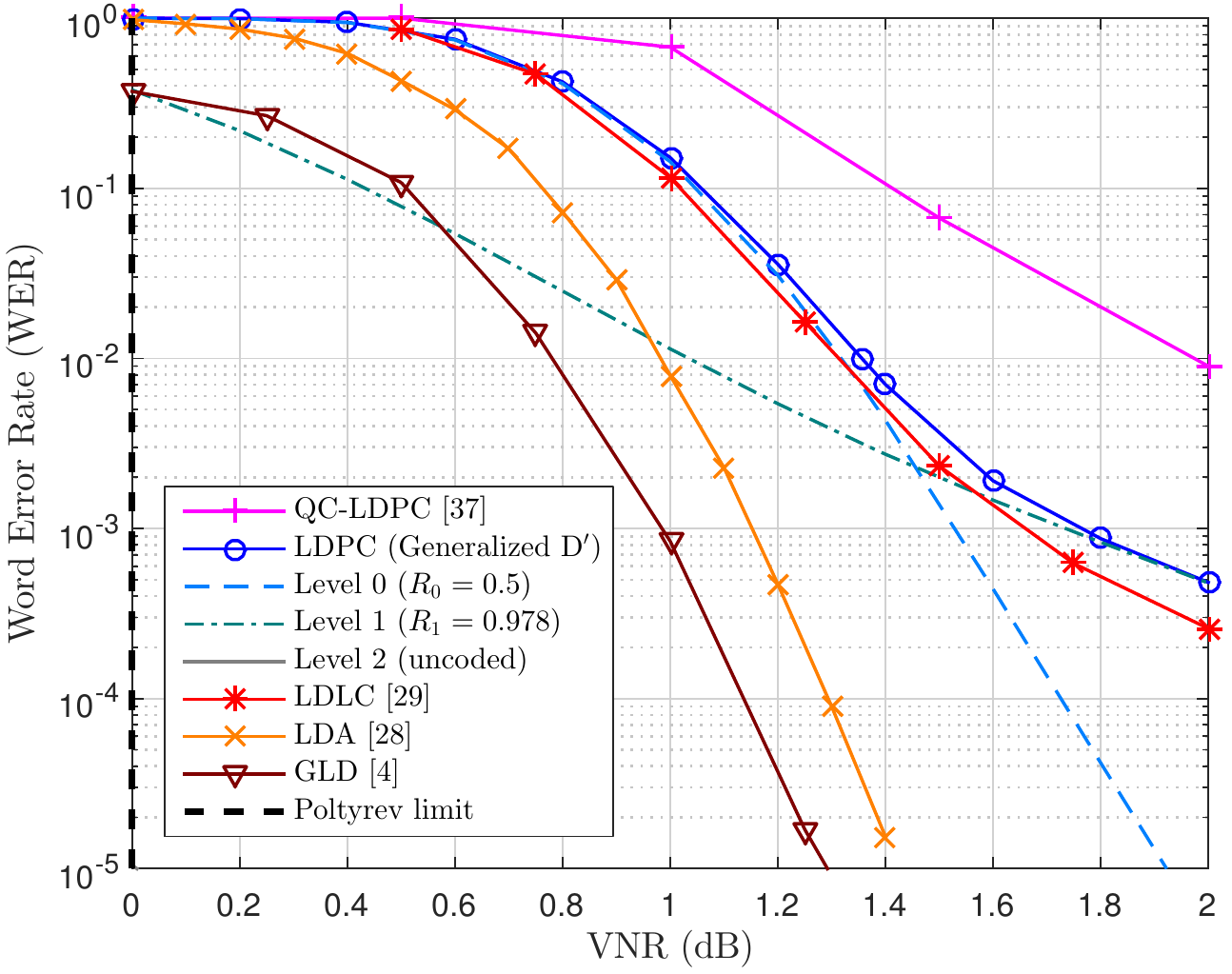}
	\end{center}
	\caption{Performance of a Generalized Construction~D$'$ 2-level LDPC lattice of dimension $1000$, designed to achieve $P_e \leq 10^{-2}$ under multistage decoding. For comparison, the performance of a 1-level QC-LDPC lattice with $n=1190$ \cite{Khodaiemehr.etal.2017.Practical-Encoder-Decoder}, an LDLC lattice with $n=1000$ \cite{Hernandez.Kurkoski.2016.Three/two-Gaussian-Parametric}, an LDA lattice with $n=1000$ \cite{Pietro.etal.2018.LDA-Lattices-Dithering}, a GLD lattice with $n=1000$ \cite{Boutros.etal.2014.Generalized-Low-density-(GLD)}, and the Poltyrev limit are also shown.}
	\label{fig:unconstrained_1e-2_n1000}
\end{figure}

Fig. \ref{fig:unconstrained_1e-2_n1000} 
shows the WER as a function of the VNR for our proposed LDPC lattice, as well as for the individual levels used in its multilevel construction. Note that the error probability of the uncoded level is so low that it does not appear in Fig. \ref{fig:unconstrained_1e-2_n1000}. It is also interesting to point out that, at the design VNR, code $\calC_0$ displays $\wer = 6.9 \cdot 10^{-3}$, whereas code $\calC_1$ displays  $\wer = 3.2 \cdot 10^{-3}$. This suggests that the optimal design criterion under multistage decoding may not be the equal error probability rule, even if only the coded levels are considered.

For the sake of comparison, Fig. \ref{fig:unconstrained_1e-2_n1000} also shows the performance of other state-of-the-art lattices with dimension around $1000$, namely: a one-level QC-LDPC lattice with $n=1190$ and rate $R = 0.786$ ($m=935$) \cite{Khodaiemehr.etal.2017.Practical-Encoder-Decoder}; an LDLC lattice with degree $7$ and $n=1000$ decoded with the three/two Gaussian parametric decoder \cite{Hernandez.Kurkoski.2016.Three/two-Gaussian-Parametric}; an LDA lattice with $n=1000$ based on an $(2,5)$-regular LDPC code over $\FF_{11}$ \cite{Pietro.etal.2018.LDA-Lattices-Dithering}; and a GLD lattice with $n=1000$ based on a $[4,3,2]$ linear code over $\FF_{11}$ \cite{Boutros.etal.2014.Generalized-Low-density-(GLD)}.

As we can see, the performance of the proposed LDPC lattice is only slightly inferior to that of the LDLC lattice, but it achieves this result with a much lower decoding complexity.

On the other hand, the QC-LDPC lattice has a much worse performance. This can be attributed to the fact that it contains a single coded level, suffering from the low performance of the uncoded level at these design parameters.

Fig. \ref{fig:unconstrained_1e-2_n1000} also shows that the LDA and GLD lattices have a significantly better performance compared to our LDPC lattice. However, these lattices rely on linear codes defined over $\FF_{11}$, leading to a much higher decoding complexity.

It can be seen in Fig. \ref{fig:unconstrained_1e-2_n1000} that our proposed LDPC lattice displays an error floor caused by low performance of $\calC_1$. This may be partly explained by the high value of $R_1$ for this block length and by the presence of a substantial amount of 4-cycles in the parity-check matrix $\bH_1$.

\subsection{Design for $P_e \leq 10^{-2}$ and $n=10000$}

For our last example, we design an LDPC lattice with dimension $n=10000$ for $P_e \leq 10^{-2}$. We use the same design rule as in \ref{ssec:1e-2_n1000}, however, we place an additional constraint on $R_1$, requiring it to be sufficiently small such that the PEG construction \cite{Hu.etal.2005.Regular-Irregular-Progressive} does not generate any 4-cycles. Following this constraint, we designed for $\wer = 10^{-2}$, arriving at rates $R_0 = 0.4094$ ($m_0 = 5906$) and $R_1 = 0.973$ ($m_1 = 270$) and $\vnr = 0.884$~dB. The simulated result indicates the crossing of $\wer = 10^{-2}$ at $\vnr = 0.8790$~dB.

\begin{figure}
	\begin{center}
		\includegraphics[width=0.5\textwidth]{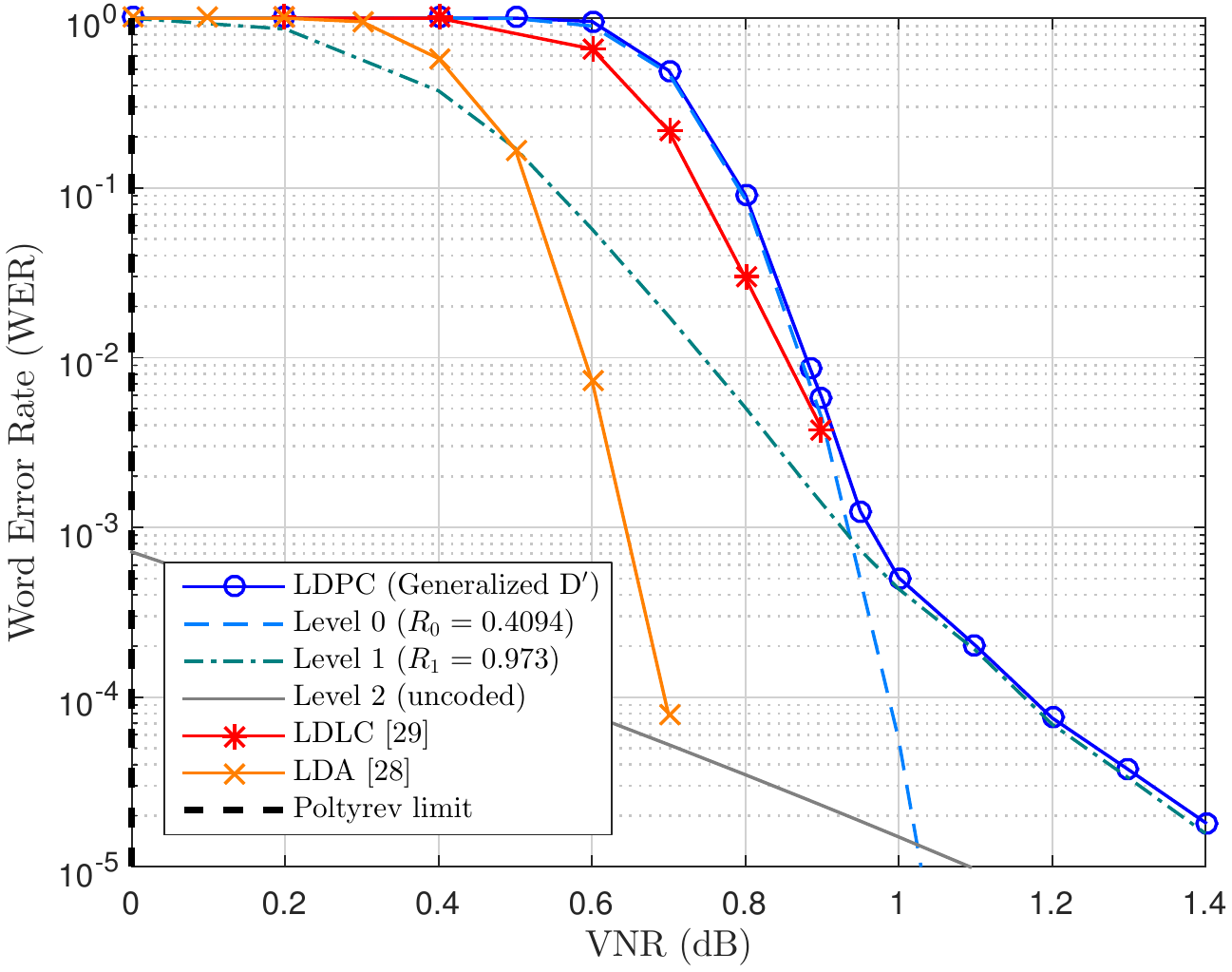}
	\end{center}
	\caption{Performance of a Generalized Construction~D$'$ 2-level LDPC lattice of dimension $10000$, designed to achieve $P_e \leq 10^{-2}$ under multistage decoding. For comparison, the performance of an LDLC lattice with $n=10000$ \cite{Hernandez.Kurkoski.2016.Three/two-Gaussian-Parametric}, an LDA lattice with $n=10000$ \cite{Pietro.etal.2018.LDA-Lattices-Dithering}, and the Poltyrev limit are also shown.}
	\label{fig:unconstrained_1e-2_n10000}
\end{figure}

Fig. \ref{fig:unconstrained_1e-2_n10000}
shows the word error rate versus VNR curve for our LDPC lattice. As benchmarks, we have also plotted performance curves for other lattices with dimension $n=10000$, including an LDLC lattice with degree $7$ and an LDA lattice, again based on a $(2,5)$-regular LDPC code over $\FF_{11}$.

Similarly to subsection \ref{ssec:1e-2_n1000}, we see that our LDPC lattice almost matches the performance of the LDLC lattice, with the benefit of less complex decoding. We can also see that the gap to the LDA performance has decreased.

\subsection{Discussion}

A limitation of the check splitting procedure is that the variable-node degree distributions for the component codes must be the same. Since the independent optimization of the component codes leads to different variable-node degree distributions, it is clear that using a single distribution for codes of significantly different rates cannot be optimal.

Note that this restriction on degree distributions is not necessarily imposed by Generalized Construction~D$'$, but is rather a consequence of check splitting. Finding a method of designing nested LDPC codes with a more flexible choice of variable-node degree distributions remains an open problem.

\section{Conclusion}
\label{sec:conclusion}

This paper has addressed the design and implementation of multilevel LDPC lattices. Two main contributions are provided, solving problems that were left open for over a decade.

First, an alternative description of Construction~D$'$ is proposed which enables sequential encoding of the component codes. As a consequence, we show that low-complexity off-the-shelf binary LDPC encoders and decoders can be adapted to produce multilevel encoding and multistage decoding algorithms for LDPC lattices with a complexity that is linear in the total number of coded bits, a property that has not been attained by any other existing Poltyrev-limit-approaching lattice.

Second, a generalization of Construction~D$'$ is proposed that relaxes the nesting constraints on the parity-check matrices of the component codes, significantly facilitating their design; specifically, under the new construction, only the component codes have to be nested, not their parity-check matrices. Following this result, we have proposed a general principle for constructing nested codes based on the partitioning of parity-check equations, as well as a practical method, inspired by the PEG algorithm, to construct LDPC subcodes of large girth that can be efficiently encoded. 

Based on this new construction, low-complexity multilevel LDPC lattices are designed whose performance under multistage decoding is shown to be comparable to that of polar lattices, closing a long-standing gap between the performances of Construction~D and Construction~D$'$ lattices. Our proposed LDPC lattices are also shown to achieve a performance close to that of LDLCs, albeit with a much lower complexity.

While the achieved performance is still far from that of $p$-ary lattices such as GLD and LDA, it should be noted that only variable-regular lattices have been considered in this work. One can reasonably expect that a much better performance may be achieved by irregular LDPC lattices with carefully designed degree distributions. However, a new design procedure is required to ensure that the resulting codes remain nested. We hope that this paper can inspire future work along this direction.

\section*{Acknowledgements}

The authors would like to thank Frank Kschischang, Amir Banihashemi and Cong Ling for helpful discussions, as well as the anonymous reviewers for their suggestions to improve the paper.

\newpage

\begin{IEEEbiographynophoto}{Paulo Ricardo Branco da Silva} received the B.Sc. degree from the Instituto Mau\'a de Tecnologia, S\~ao Caetano do Sul, Brazil, in 2012, and the M.Sc. degree from the Federal University of Santa Catarina, Florian\'opolis, Brazil, in 2015, all in electrical engineering. He is currently a doctoral student in electrical engineering at the latter university. His main research interests lie in channel coding, information theory, and network coding. He is a member of the Brazilian Telecommunications Society (SBrT).
\end{IEEEbiographynophoto}

\begin{IEEEbiographynophoto}{Danilo Silva} (S'06--M'09) received the B.Sc.\ degree from the Federal University of Pernambuco (UFPE), Recife, Brazil, in 2002, the M.Sc.\ degree from the Pontifical Catholic University of Rio de Janeiro (PUC-Rio), Rio de Janeiro, Brazil, in 2005, and the Ph.D. degree from the University of Toronto, Toronto, Canada, in 2009, all in electrical engineering.

From 2009 to 2010, he was a Postdoctoral Fellow at the University of Toronto, at the \'Ecole Polytechnique F\'ed\'erale de Lausanne (EPFL), and at the State University of Campinas (UNICAMP).
In 2010, he joined the Department of Electrical and Electronic Engineering, Federal University of Santa Catarina (UFSC), Brazil, where he is currently an Associate Professor. His research interests include wireless communications, channel coding, information theory, and machine learning.

Dr. Silva is a member of the Brazilian Telecommunications Society (SBrT). He was a recipient of a CAPES Ph.D. Scholarship in 2005, the Shahid U. H. Qureshi Memorial Scholarship in 2009, and a FAPESP Postdoctoral Scholarship in 2010.
\end{IEEEbiographynophoto}

\enlargethispage{-5.0in}

\end{document}